\documentclass[useAMS,usenatbib]{mn2e}

\usepackage{epsfig}
\usepackage{amsfonts}
\usepackage[usenames]{color}
\usepackage[usenames,dvipsnames]{xcolor}
\usepackage{url}
\usepackage{subfigure}
\usepackage{times,graphicx,amsmath,amsfonts,amssymb,aas_macros,epstopdf,hyperref}
\usepackage[normalem]{ulem}
\usepackage[T1]{fontenc}
\usepackage{multirow}
\newcommand{\angstrom}{\textup{\AA}}

\usepackage{hyperref}
\hypersetup{
    colorlinks=true,
    linkcolor=blue,
    citecolor=blue,
}

\def\etal{{\frenchspacing\it et al.}}
\def\ie{{\frenchspacing\it i.e.}}
\def\eg{{\frenchspacing\it e.g.}}

\def\be{\begin{equation}}
\def\ee{\end{equation}}
\def\ba{\begin{eqnarray}}
\def\ea{\end{eqnarray}}

\def\nn{\nonumber}

\newcommand{\hompc}{\,h\,{\rm Mpc}^{-1}}
\newcommand{\mpcoh}{\,h^{-1}\,{\rm Mpc}}

\def\vunits{{\,h^3{\rm Mpc}^{-3}}}

\def\d{{\rm d}}
\def\rr{{\bf r}}

\def\k{{\bf k}}

\def\LaTeX{L\kern-.36em\raise.3ex\hbox{a}\kern-.15em
    T\kern-.1667em\lower.7ex\hbox{E}\kern-.125emX}

\begin{document}

\voffset-1.25cm
\title[Tomographic BAO analysis in Fourier space]{The clustering of galaxies in the completed SDSS-III Baryon Oscillation Spectroscopic Survey: tomographic BAO analysis of DR12 combined sample in Fourier space}
\author[Zhao \etal]{
\parbox{\textwidth}{
Gong-Bo Zhao$^{1,2}$\thanks{Email: \url{gbzhao@nao.cas.cn}}, Yuting Wang$^{1,2}$, Shun Saito$^{3,4}$, Dandan Wang$^{1}$, Ashley J. Ross$^{5,2}$, Florian Beutler$^{2,6}$, Jan Niklas Grieb$^{7,8}$, Chia-Hsun Chuang$^{9,10}$, Francisco-Shu Kitaura$^{10}$, Sergio Rodriguez-Torres$^{9,11,12}$, Will J. Percival$^{2}$, Joel  R. Brownstein$^{13}$, Antonio J. Cuesta$^{14}$, Daniel J. Eisenstein$^{15}$, H\'ector Gil-Mar\'{\i}n$^{16,17,2}$, Jean-Paul Kneib$^{18}$, Robert C. Nichol$^{2}$,  Matthew D. Olmstead$^{19}$, Francisco Prada$^{9}$, Graziano Rossi$^{20}$, Salvador Salazar-Albornoz$^{21,22}$, Lado Samushia$^{2}$, Ariel G. S\'anchez$^{22}$, Daniel Thomas$^{2}$, Jeremy L. Tinker$^{23}$, Rita Tojeiro$^{24}$, David H. Weinberg$^{25,5}$, Fangzhou Zhu$^{26}$}
\vspace*{15pt} \\
$^{1}$ National Astronomy Observatories, Chinese Academy of Science, Beijing, 100012, P.R.China\\
$^{2}$ Institute of Cosmology \& Gravitation, University of Portsmouth, Dennis Sciama Building, Portsmouth, PO1 3FX, UK\\
$^{3}$ Max-Planck-Institut f\"ur Astrophysik, Karl-Schwarzschild-Star{\ss}e 1, D-85740 Garching bei M\"unchen, Germany\\
$^{4}$ Kavli Institute for the Physics and Mathematics of the Universe (IPMU), The University of Tokyo, Kashiwa, Chiba 277-8583, Japan \\
$^{5}$ Center for Cosmology and AstroParticle Physics, The Ohio State University, Columbus, OH 43210, USA \\
$^{6}$ Lawrence Berkeley National Lab, 1 Cyclotron Rd, Berkeley CA 94720, USA\\
$^{7}$ Max-Planck-Institut f\"ur extraterrestrische Physik, Postfach 1312, Giessenbachstr., 85741 Garching, Germany\\
$^{8}$ Universit\"ats-Sternwarte M\"unchen, Ludwig-Maximilians-Universit\"at M\"unchen, Scheinerstra{\ss}e 1, 81679 M\"unchen, Germany\\
$^{9}$ Instituto de F\'{\i}sica Te\'orica, (UAM/CSIC), Universidad Aut\'onoma de Madrid,  Cantoblanco, E-28049 Madrid, Spain \\
$^{10}$ Leibniz-Institut f\"ur Astrophysik Potsdam (AIP), An der Sternwarte 16, 14482 Potsdam, Germany \\
$^{11}$ Campus of International Excellence UAM+CSIC, Cantoblanco, E-28049 Madrid, Spain \\
$^{12}$ Departamento de F\'{\i}sica Te\'orica, Universidad Aut\'onoma de Madrid, Cantoblanco, E-28049, Madrid, Spain\\
$^{13}$ Department of Physics and Astronomy, University of Utah, 115 S 1400 E, Salt Lake City, UT 84112, USA \\
$^{14}$ Institut de Ci\`encies del Cosmos (ICCUB), Universitat de Barcelona (IEEC- UB), Mart\'{\i} i Franqu\`es 1, E-08028 Barcelona, Spain \\
$^{15}$ Harvard-Smithsonian Center for Astrophysics, 60 Garden St., Cambridge, MA 02138, USA \\
$^{16}$ Sorbonne Universit\'es, Institut Lagrange de Paris (ILP), 98 bis Boulevard Arago, 75014 Paris, France\\
$^{17}$ Laboratoire de Physique Nucl\'eaire et de Hautes Energies, Universit\'e Pierre et Marie Curie, 4 Place Jussieu, 75005 Paris, France\\
$^{18}$ Laboratoire d'Astrophysique, Ecole Polytechnique F\'ed\'erale de Lausanne (EPFL), Observatoire de Sauverny, CH-1290 Versoix, Switzerland\\
$^{19}$ Department of Chemistry and Physics, King's College, 133 North River St, Wilkes Barre, PA 18711, USA \\
$^{20}$ Department of Astronomy and Space Science, Sejong University, Seoul 143-747, Korea \\
$^{21}$ Universit\"ats-Sternwarte M\"unchen, Ludwig-Maximilians-Universit\"at Munchen, Scheinerstra{\ss}e 1, 81679 M\"unchen, Germany\\
$^{22}$ Max-Planck-Institut f\"ur extraterrestrische Physik, Postfach 1312, Giessenbachstr., 85741 Garching, Germany\\
$^{23}$ Center for Cosmology and Particle Physics, Department of Physics, New York University, 4 Washington Place, New York, NY 10003, USA\\
$^{24}$ School of Physics and Astronomy, University of St Andrews, North Haugh, St Andrews KY16 9SS, UK\\
$^{25}$ Department of Astronomy, The Ohio State University, 140 West 18th Avenue, Columbus, OH 43210, USA\\
$^{26}$ Department of Physics, Yale University, New Haven, CT 06511, USA
}
\date{\today} 
\pagerange{\pageref{firstpage}--\pageref{lastpage}}

\label{firstpage}

\maketitle

\clearpage

\begin{abstract} 

We perform a tomographic baryon acoustic oscillations (BAO) analysis using the monopole, quadrupole and hexadecapole of the redshift-space galaxy power spectrum measured from the pre-reconstructed combined galaxy sample of the completed Sloan Digital Sky Survey (SDSS-III) Baryon Oscillation Spectroscopic Survey (BOSS) Data Release (DR)12 covering the redshift range of $0.20<z<0.75$. By allowing for overlap between neighbouring redshift slices, we successfully obtained the isotropic and anisotropic BAO distance measurements within nine redshift slices to a precision of $1.5\%-3.4\%$ for $D_V/r_d$, $1.8\% -4.2\%$ for $D_A/r_d$ and $3.7\% - 7.5\%$ for $H \ r_d$, depending on effective redshifts. We provide our BAO measurement of  $D_A/r_d$ and $H \ r_d$ with the full covariance matrix, which can be used for cosmological implications. Our measurements are consistent with those presented in \citet{Acacia}, in which the BAO distances are measured at three effective redshifts. We constrain dark energy parameters using our measurements, and find an improvement of the Figure-of-Merit of dark energy in general due to the temporal BAO information resolved. This paper is part of a set that analyses the final galaxy clustering dataset from BOSS.
\end{abstract}


\section{Introduction}
\label{sec:intro}

One of key science drivers of large spectroscopic galaxy surveys is to unveil the nature of dark energy (DE), the unknown energy component with a negative pressure to drive the accelerating expansion of the Universe \citep{Riess,Perlmutter}. The equation-of-state (EoS) function $w(z)$, which is the ratio of pressure over energy density of DE and is a function of redshift $z$ in general, is a proxy linking the nature of DE and its phenomenological features which can be probed by observations. For instance, a observational confirmation of $w=-1$ may suggest that DE is essentially vacuum energy, while a time-evolving $w$ can be a sign of new physics, \eg, dynamical dark energy scenarios \citep{quintessence1,quintessence2,phantom,quintom,kessence}, or a breakdown of general relativity on cosmological scales (see \citealt{MGreview} for a recent review of modified gravity theories). Therefore reconstructing $w(z)$ directly from data is an efficient way for DE studies \citep{DEreview,Zhao12,DErecon06}.    

The function $w(z)$ of the DE equation of state leaves imprints on the cosmic background expansion history, which can be probed by the effect of baryon acoustic oscillations (BAO) measured from galaxy surveys \citep{Eisenstein05,Cole2005}, besides other probes including supernovae Type Ia (SN Ia) \citep{Riess,Perlmutter}, cosmic microwave background (CMB) \citep{planck15} and so forth. BAO is a characteristic three-dimensional clustering pattern of galaxies at about 150 Mpc on the comoving scale, due to sound waves generated by the photon-baryon coupling in the early universe \citep{PY70,SZ70,EH98}. The BAO distance is traditionally measured using two-point correlation functions or power spectrum of galaxies. Recent studies find that higher-order statistics of galaxies \citep{BAO3pt}, or two-point clustering of voids can also be used for BAO measurements \citep{voidBAO}\footnote{In this work, we focus on galaxies as cosmic tracers thus will only refer to galaxies when discussing BAO measurements.}. Since the BAO scale is sensitive to cosmic geometry and it is largely immune to systematics \citep{imsys}, BAO is widely used as the `standard ruler' to calibrate the expansion rate of the Universe.

Under assumptions that the BAO scale is the same in all directions with respect to the line-of-sight (l.o.s.) of the observer, one can probe the isotropic, one-dimensional (1D) BAO scale $D_V(z)\equiv \left[ cz (1+z)^2 D_A(z)^2 H^{-1}(z) \right]^{1/3}$, where $D_A(z)$ and $H(z)$ are the angular diameter distance and Hubble parameter at an effective redshift $z$ of the galaxy sample, using the monopole of the correlation function, or power spectrum of galaxies in redshift space.  

In fact, $D_A(z)$ and $H(z)$ can be separately measured when higher-order multipoles, \eg, the quadrupole and hexadecapole, are included in the analysis. This is due to the Alcock-Paczynski (AP) effect \citep{AP}: if one uses a wrong cosmology to convert redshifts into distances for the clustering analysis, the scales along and cross the l.o.s. will be dilated differently, which produces a measurable effect to break the degeneracy between $D_A$ and $H$ in the anisotropic, two-dimensional (2D) BAO analysis. The 2D BAO distances are more challenging to measure, but it is much more informative for DE studies because $w(z)$ is closely related to the first derivative of $H(z)$.

The 1D and 2D BAO signals have been detected by a number of large galaxy surveys including the Sloan Digital Sky Survey (SDSS) \citep{Eisenstein05,Percival10, DR9,BAODR11,Gil-Mar,Cuesta,Acacia,Beutler16a,Beutler16b,Ross16}, the 2-degree Field Galaxy Redshift Survey (2dFGRS) \citep{Cole2005}, WiggleZ \citep{wigglez,2012arXiv1210.2130P}, the 6-degree Field Galaxy Survey (6dFGS) \citep{6dF} and so on. The Baryon Oscillation Spectroscopic Survey (BOSS) \citep{Dawson12}, part of SDSS III project \citep{SDSS3}, has reached percent level BAO measurements at $z_{\rm eff}=0.32$ and $z_{\rm eff}=0.57$ \citep{BAODR11,Gil-Mar,Cuesta,Beutler16a,Ross16} using the `low-redshift' (LOWZ; $0.15<z<0.43$) and `constant stellar mass' samples (CMASS; $0.43<z<0.7$) of Data Release (DR) 12 \citep{Alam} \footnote{The DR12 dataset is publicly available at \url{http://www.sdss.org/dr12/}}. 

It is true that using galaxies across wide redshift ranges can yield a precise BAO measurement at a single effective redshift, but this does not capture the tomographic information in redshift, which is required for the study of $w(z)$. Subdividing the galaxy sample into a small number of  independent redshift slices and perform the BAO analysis in each slice can in principle recover the temporal information to some extent (see \citealt{Acacia} and \citealt{CF_4bins} for a three-bin and four-bin BAO analysis of the BOSS DR12 sample respectively). However, as the slice number increases, galaxies in each slice decrease, and we are at a risk of ending up with a seriously biased measurement due to large systematic uncertainties.

One possible solution is to perform the BAO analysis in overlapping redshift slices. This on one hand guarantees the sufficiency of galaxy numbers in each subsample, on the other hand, it allows for a higher temporal resolution. In this work, we perform such a tomographic BAO analysis in Fourier space using the DR12 galaxy sample, and quantify the gain in dark energy studies.     

This paper is structured as follows. In Section 2, we describe the BOSS DR12 galaxy catalogues used for our analysis, and in Section 3, we perform a Fisher matrix forecast on this sample to determine the redshift binning, and present the power spectrum measurements. We perform the BAO analysis in Section 4, and apply our measurement to dark energy studies in Section 5, before we conclude in Section 6.   

\section{The BOSS DR12 Combined Sample}
\label{sec:data}

\begin{figure}
\centering
{\includegraphics[scale=0.32]{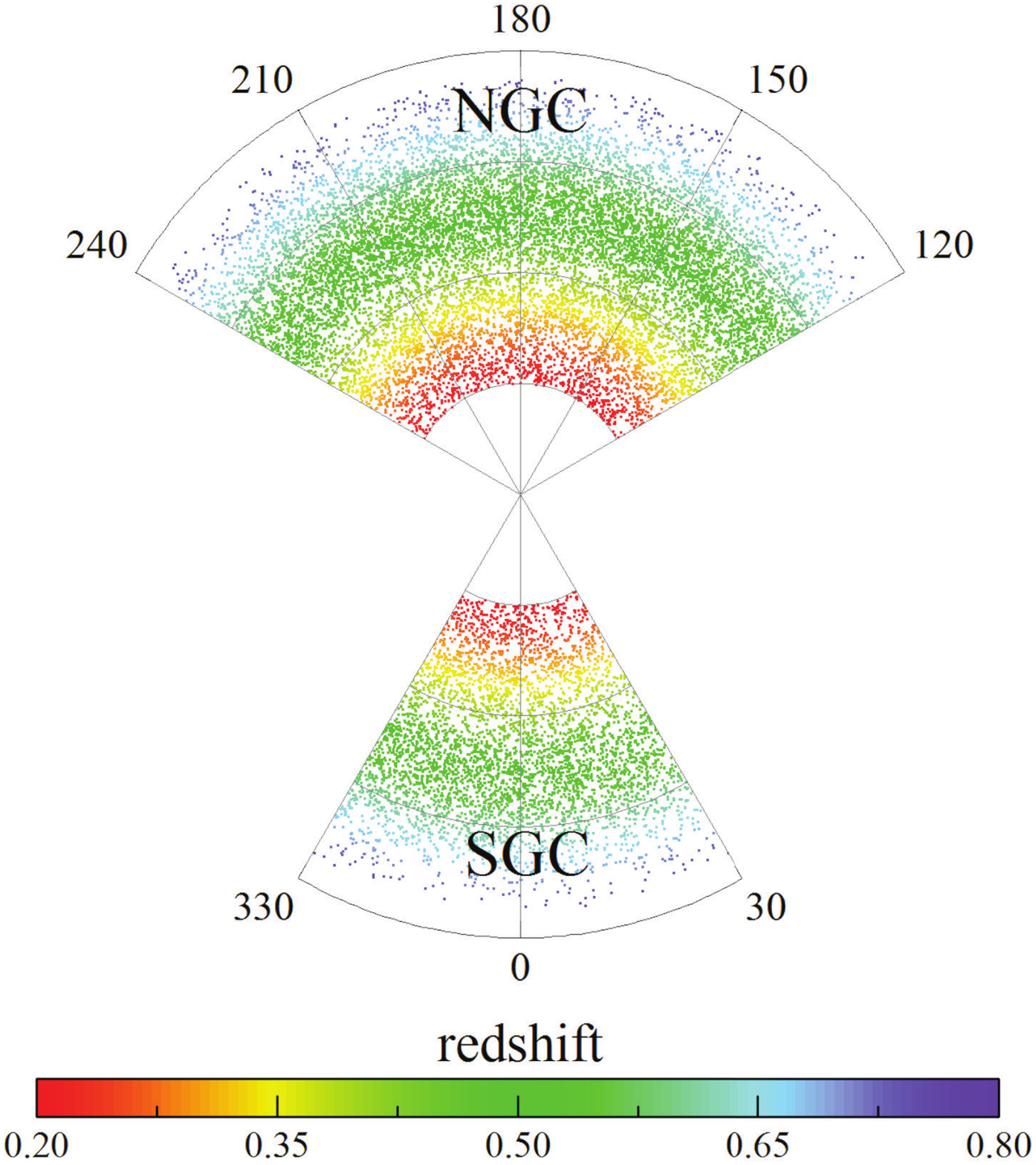}}
\caption{A wedge plot of the DR12 galaxies in the redshift range of $0.2<z<0.75$ in the NGC (upper part) and the SGC (lower part).}
\label{fig:footprint}
\end{figure}

 \begin{figure}
\centering
{\includegraphics[scale=0.32]{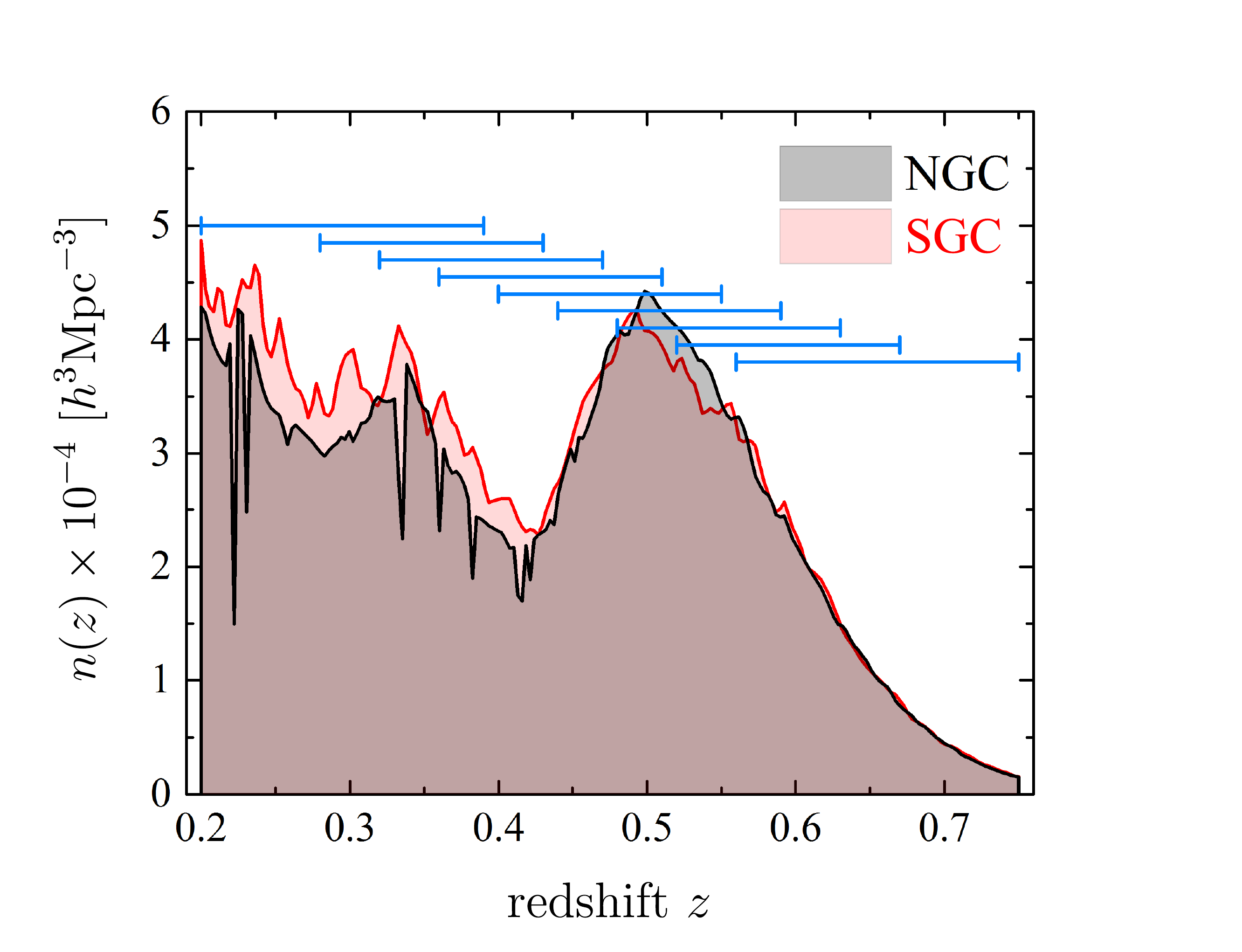}}
\caption{The volume number density of galaxies in units of $\vunits$ in the NGC (black shaded) and SGC (red shaded). The nine horizontal overlapping bins illustrate our binning scheme.}
\label{fig:nz}
\end{figure}

The BOSS program covers near 10, 000 square degrees of the sky using a $2.5$ metre-aperture Sloan Foundation Telescope \citep{SDSStelescope} at the Apache Point Observatory (APO) in New Mexico. The BOSS team has obtained spectra of more than 1.5 million galaxies brighter than $i=19.9$ and approximately 170, 000 new quasars with redshifts $2.1\leq z \leq 3.5$ to a depth of $g<22$ using the improved double-armed spectrographs in a wavelength range of $3, 600 \ \angstrom <\lambda<10, 000 \ \angstrom$. The filter, spectrograph and pipeline of BOSS are described in \citet{ugrizFilter,BOSSpipeline,BOSSspectro}. 

The DR12 combined sample is a coherent combination of two distinct targets, LOWZ and CMASS. We refer the stellar-mass incompleteness of the LOWZ and CMASS samples to \citet{Leauthaud16} and its impact on the clustering to \citet{Saito16} and \citet{Rodriguez-Torres16}. The DR12 combined catalogue is created from the observational data using the pipeline described in \citet{Reid16}, in which the survey footprint, veto masks and survey systematics are taken into account to produce the data and random catalogues. The redshift range of this sample is $0.2<z<0.75$, and it contains $\sim865, 000$ and $\sim330, 000$ galaxies in the North Galactic Cap (NGC) ($\sim5900$ deg$^2$) and South Galactic Cap (SGC) ($\sim2500$ deg$^2$) respectively. The wedge plot (Fig. \ref{fig:footprint}) visualises the DR12 sample \footnote{For the purpose of visualisation, only 2\% randomly selected galaxies are included in Fig. \ref{fig:footprint}.}. The redshift distribution of the galaxies in the NGC and SGC is shown in Fig. \ref{fig:nz}. We refer the readers to Table 2 of \citet{Reid16} for more details of the DR12 combined sample.

For each galaxy in the data catalogue, the following information is provided: the right ascension (RA), declination (DEC), redshift $z$ and a set of weights including a FKP weight \citep{FKP} $w_{\rm FKP}$, which is crucial to optimise the signal-to-noise ratio of power spectrum measurements, a systematic weight, $w_{\rm sys}$ to account for systematic effects from the contamination of stars and variations in seeing conditions, a redshift failure weight, $w_{\rm rf}$ to avoid using the galaxy without a robust redshift estimate, and a fibre collision weight, $w_{\rm fc}$ to correct for the clustering signal on small scales due to the fibre collision. With all the weights accounted for, each individual galaxy is counted as an effective number of,
\be\label{eq:weight} w_{\rm T}=w_{\rm FKP}w_{\rm c}; \ w_{\rm c}=w_{\rm sys}(w_{\rm fc}+w_{\rm rf}-1)\ee More details of the weights are described in \citet{imsys,BAODR11,Ross16}.

\begin{table*}
\begin{center}
\begin{tabular}{ccccccccc}
\hline\hline
   redshift bin index &  redshift range & effective $z$& $N_{\rm NGC}$ & $N_{\rm SGC}$ & $N_{\rm tot}$& $\sigma_{D_A}/D_A $  & $\sigma_{H}/H $  &$\sigma_{D_V}/D_V$   \\ \hline

$z$ bin 1	&	$0.20	<	z	<	0.39$	&	0.31 &	176,899	&	75,558  &		252,457	&	0.029	&	0.0705 &	0.024 \\ 
$z$ bin 2	&	$0.28	<	z	<	0.43$	&	0.36	&	194,754	&	81,539 &		276,293        &	0.028	&	0.0681 &	0.023\\
$z$ bin 3	&	$0.32	<	z	<	0.47$	&	0.40 &	230,388	&	93,825 &		324,213	&	0.025	&	0.0616 &	0.021\\
$z$ bin 4	&	$0.36	<	z	<	0.51$	&	0.44	&	294,749	&     115,029 &		409,778	&	0.023	&	0.0553 &	0.018\\
$z$ bin 5	&	$0.40	<	z	<	0.55$	&	0.48 &	370,429	&     136,117 &		 506,546       &	0.020	&	0.0502 &	0.017\\
$z$ bin 6	&	$0.44	<	z	<	0.59$	&	0.52 &	423,716	&      154,486 &		578,202		&	0.019	&	0.0464 &	0.016\\
$z$ bin 7	&	$0.48        <	z	<	0.63$	&	0.56 &	410,324	&	149,364 &		559,688		&	0.018	&	0.0441 &	0.015\\
$z$ bin 8	&	$0.52        <	z	<	0.67$	&	0.59 &	331,067	&	121,145 &		452,212		&	0.018	&	0.0436 &	0.015\\
$z$ bin 9	&	$0.56        <	z	<	0.75$	&	0.64 &	231,505	&	86,576  &		318,081		&	0.019	&	0.0418 &	0.014\\

\hline\hline
\end{tabular}
\end{center}
\caption{Statistics of the galaxies within nine overlapping redshift bins, and the corresponding Fisher forecast result for the BAO parameters.}
\label{tab:zbins}
\end{table*}

\begin{figure*}
\centering
{\includegraphics[scale=0.3]{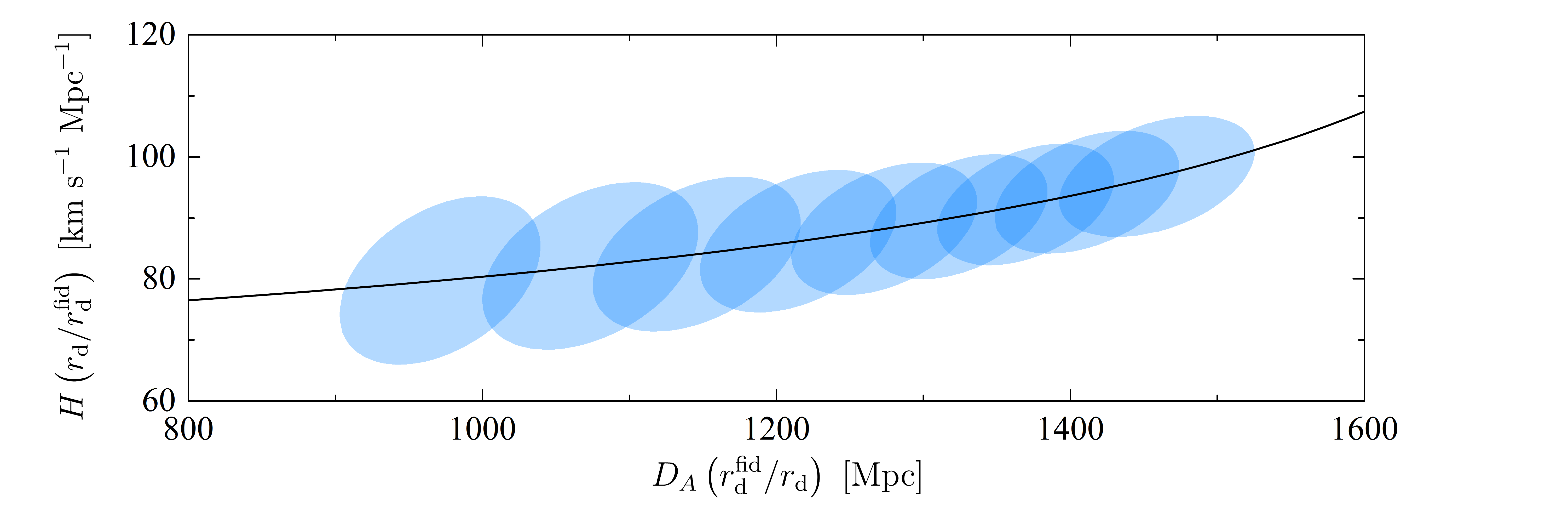}}
\caption{The 95\% CL contour plots for $D_A(r_{\rm d}^{\rm fid}/r_{\rm d})$ and $H(r_{\rm d}/r_{\rm d}^{\rm fid})$ derived from a Fisher matrix forecast for DR12 galaxies in nine redshift slices. For contours from left to right, the effective redshifts of galaxies used increase from $z_{\rm eff}=0.31$ to $z_{\rm eff}=0.64$. The black solid curve shows the prediction of the fiducial model used in this analysis.}
\label{fig:DA_H_Fisher}
\end{figure*}

For the clustering analysis, the auxiliary random catalogues, whose spatial and redshift distributions match those of data catalogues but without any clustering structure, are required. In this analysis, we use the random catalogues consisting of $50$ times the number of galaxies to reduce the sampling noise. 

Since we only observe one realisation of the distribution of galaxies in the past lightcone, we need a large number of additional realisations, which can be obtained using numerical simulations, to estimate the data covariance matrix. In this work, we use the MultiDark PATCHY (MD-Patchy) mock catalogues \citep{patchy}, which provide $2048$ realisations of galaxy distribution matching the spatial and redshift distributions of the DR12 data sample. These galaxy mocks can accurately recover the input two-point and three-point statistics, and are sufficient for the calibration of data covariance matrix for this analysis. 

Given RA, DEC and $z$, the Cartesian coordinates of the galaxy concerned, the distances between galaxy pairs can be calculated, given a fiducial cosmology. It is true that the final BAO measurement is independent of the fiducial cosmology used for the redshift-distance conversion, and in the production of mock catalogues, we choose the same cosmological parameters used in the MD-Patchy mocks for convenience, \ie, \be\label{eq:fid}\{\Omega_M, \Omega_b, \Omega_K, h, \sigma_8\} = \{0.307115, 0.0480, 0, 0.6777, 0.8288 \}\ee which is consistent with the results from the Planck collaboration \citep{planck15} \footnote{This publication will be referred to as `Planck 2015' in later texts.}. 

\section{Tomographic BAO measurements}

\subsection{Preparations}
\label{sec:forecasts}

To ensure a robust BAO distance measurement within each redshift slice while maximising the tomographic information, we employ a Fisher matrix forecast following the method developed in \citet{BAOFisher}. Using the $k$ modes up to $0.3\hompc$ for the BAO analysis without the reconstruction process \citep{BAOrecon}, we require that the precision of the isotropic BAO distance measurement within each redshift bin is better than 3\%, while for the anisotropic BAO measurement, the precision on $D_A$ and $H$ within each bin is no worse than 4\% and 8\% respectively. This is roughly the BAO sensitivity of the SDSS-II DR7 sample \citep{Percival10}. We allow for the overlapping between neighbouring redshift bins to be 75\% maximal, which well balances the redshift resolution and the complementarity of the information between overlapping bins \footnote{Based on a Fisher matrix analysis, we find that allowing for more overlap between neighbouring bins does not further improve the FoM of dark energy, which means that the BAO information extracted from our binning scheme saturates. On the other hand, a high level overlap among bins can yield a singular data covariance matrix, which is problematic for the likelihood analysis.}. This yields a binning scheme visualised in Fig \ref{fig:nz} and in Table \ref{tab:zbins}. As shown, the entire sample is subdivided into nine bins, with the maximal overlapping to be 73\%. The BAO projection result in Table \ref{tab:zbins} satisfies the requirement mentioned above, \ie, the worst isotropic and anisotropic BAO distance measurement is predicted to be 2.4\% ($D_V$), 2.9\% ($D_A$) and 7\% ($H$) respectively. The predicted 68 and 95\% confidence level (CL) contours between $D_A$ and $H$ using galaxies in nine bins are shown in Fig. \ref{fig:DA_H_Fisher}. The black solid curve illustrates the fiducial model.    

\subsection{The interpolation scheme}

\begin{figure*}
\centering
{\includegraphics[scale=0.45]{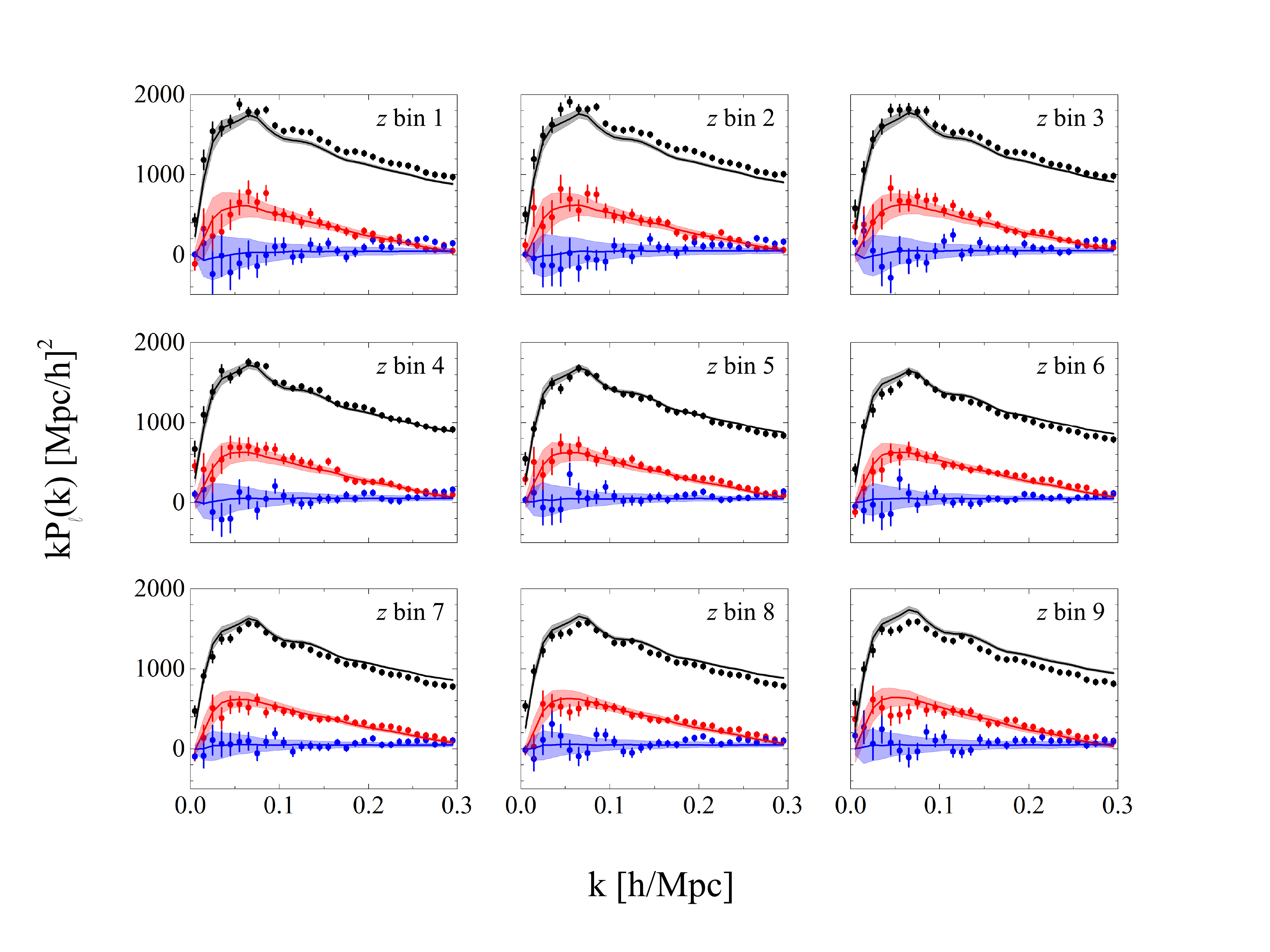}}
\caption{The power spectrum monopole (black), quadrupole (red) and hexadecapole (blue) measurements from the galaxy catalogue (data points) and from the MD-Patchy mock catalogue (shaded region) for the NGC. The solid curves are the average of all the mocks, and the error bars and error bands show the standard deviation at each $k$ bin.}
\label{fig:pk_ngc}
\end{figure*}

\begin{figure*}
\centering
{\includegraphics[scale=0.45]{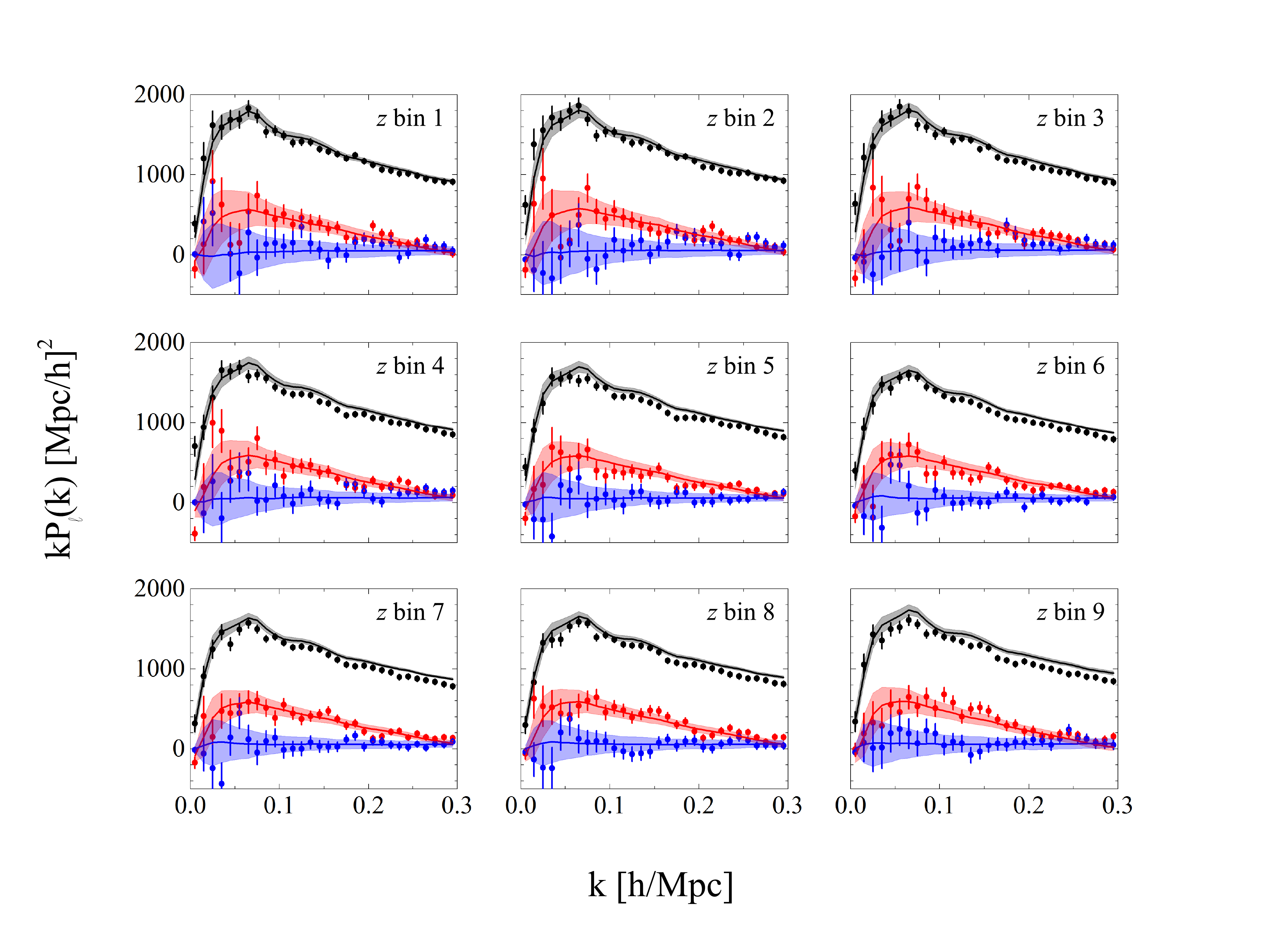}}
\caption{Same as Fig \ref{fig:pk_ngc} but for the SGC.}
\label{fig:pk_sgc}
\end{figure*}

The first step for the power spectrum multipole measurement is to assign the galaxies and randoms to a regular Cartesian grid, and choose an interpolation scheme to obtain a smoothed overdensity field for the Fourier analysis in subsequent steps. 

In this work, we embed the entire survey volume into a cubic box with $L =  5000\mpcoh$ a side \footnote{We have tested and found that a box with this size is sufficiently large to cover the entire survey volume.}, and the box is subdivided into $N_g^3=1024^3$ cubic cells. To obtain the smoothed overdensity field, an interpolation scheme is needed for the mass assignment. It is well known that the aliasing problem is inevitable in Fourier analysis, but choosing a suitable interpolation scheme (with corrections after the Fourier transformation; see discussions later) can largely reduce the aliasing to a negligible level at the scale for the BAO analysis.

Traditional interpolation schemes include the Nearest-Grid-Point (NGP), Cloud-in-Cell (CIC), Triangular-Shaped-Cloud (TSC) and so on. These correspond to the first, second and third order B-spline interpolations. The higher order it is, the less level of aliasing survives after the correction \citep{interlacing, Jing05}.  

Recently, \citet{interlacing} found that using the fourth-order B-spline, also called the Piecewise Cubic Spline (PCS) interpolation, can suppress the aliasing effect to a level below $0.1\%$ even at the Nyquist scale after the correction. In this work, we follow \citet{interlacing} and use the PCS interpolation to calculate the overdensity field on the grid, which is equivalent to convolving the underlying overdensity field with the following window function $W_{\rho}(s)$ in configuration space, \ba 
W_{\rho}(s)=
\left\{
\begin{array}{cc}
 \frac{1}{6} (4-6s^2+3|s|^3) & 0\leqslant|s|<1     \\
 \frac{1}{6} (2-|s|)^3   & 1\leqslant|s|<2     \\
 0 & {\rm otherwise  }      
\end{array}
\right.
\ea where $s$ denotes the separation between grids (in unit of number of grids) in one dimension. This means that the overdensity in each cell is contributed by galaxies and randoms in its $N_{\rm PCS}=5^3$ neighbouring cells \footnote{For a reference, $N_{\rm NGP}=1; N_{\rm CIC}=2^3, N_{\rm TSC}=3^3$.}. 

After the interpolation, we obtain an overdensity field $\Delta(\rr)$, \be\label{eq:Delta} \Delta(\rr)\equiv\frac{w_{\rm T}(\rr)}{\sqrt{N}}[n_{\rm G}(\rr)-\gamma n_{\rm R}(\rr)]\ee where $N$ is a normalisation factor which can be computed using the random catalogue \citep{FKP}, \be\label{eq:norm} N = \gamma \sum_{i=1}^{N_{\rm R}} n_{\rm G}(\rr_i) w_{\rm FKP}^2(\rr_i)\ee The summation here is over $N_{\rm R}$ samples in the random catalogue. The quantity $w_{\rm T}$ is the total weight for the concerning galaxy given in Eq (\ref{eq:weight}), $n_{\rm G}$ and $n_{\rm R}$ are the number density at position $\rr$ of the galaxy and random catalogues respectively, $\gamma$ is the ratio between the total sample numbers of the galaxy ($N_{\rm G}$) and random ($N_{\rm R}$) catalogues, \ie, $\gamma = N_{\rm G}/N_{\rm R}$ and in this work $\gamma\sim0.02$.

\subsection{The estimator for $P_{\ell}(k)$}

\begin{figure*}
\centering
{\includegraphics[scale=0.35]{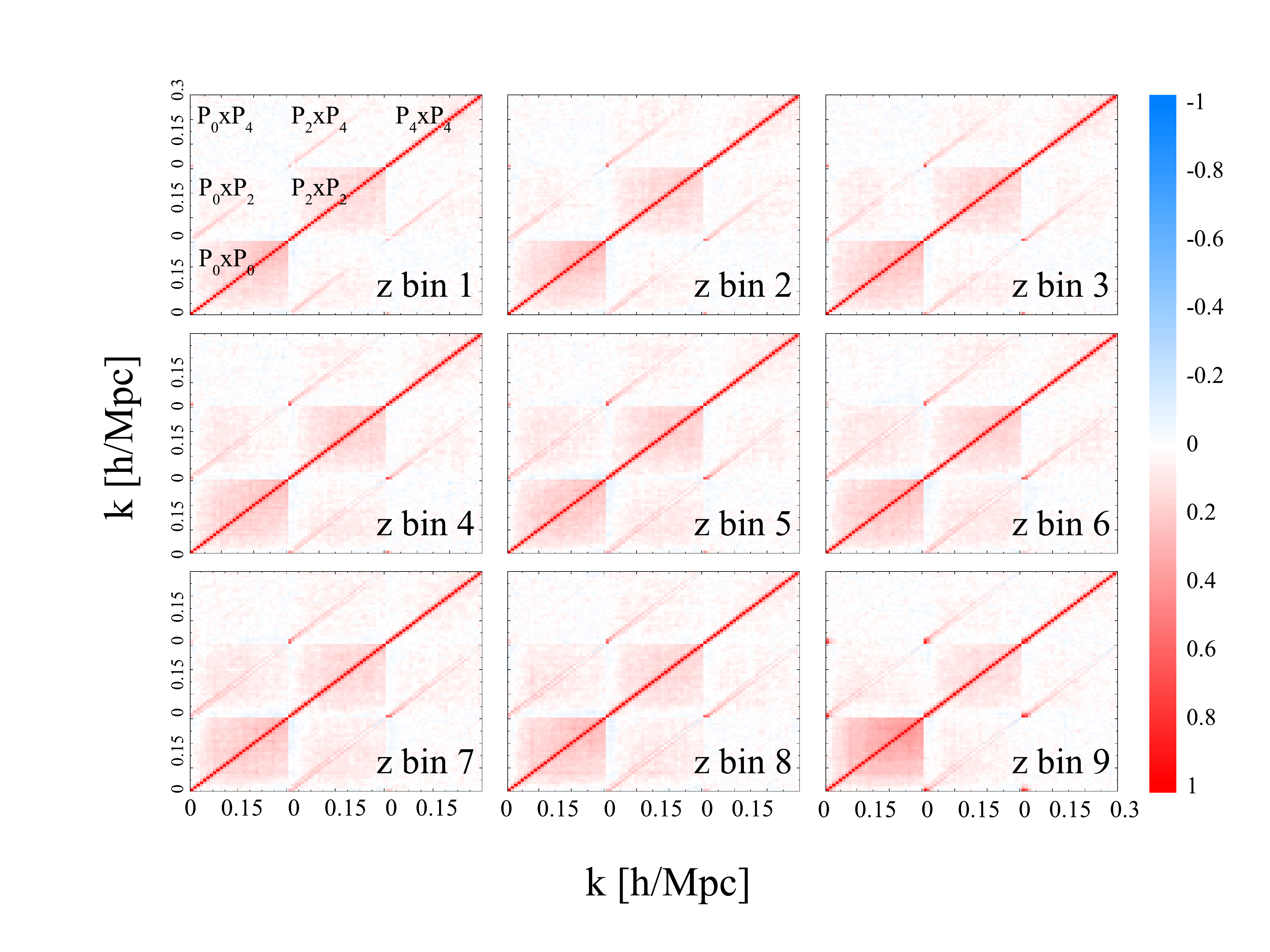}}
\caption{The correlation matrix of $P_{\ell}(k)$ for the galaxies in the NGC.}
\label{fig:corr_ngc}
\end{figure*}

\begin{figure*}
\centering
{\includegraphics[scale=0.35]{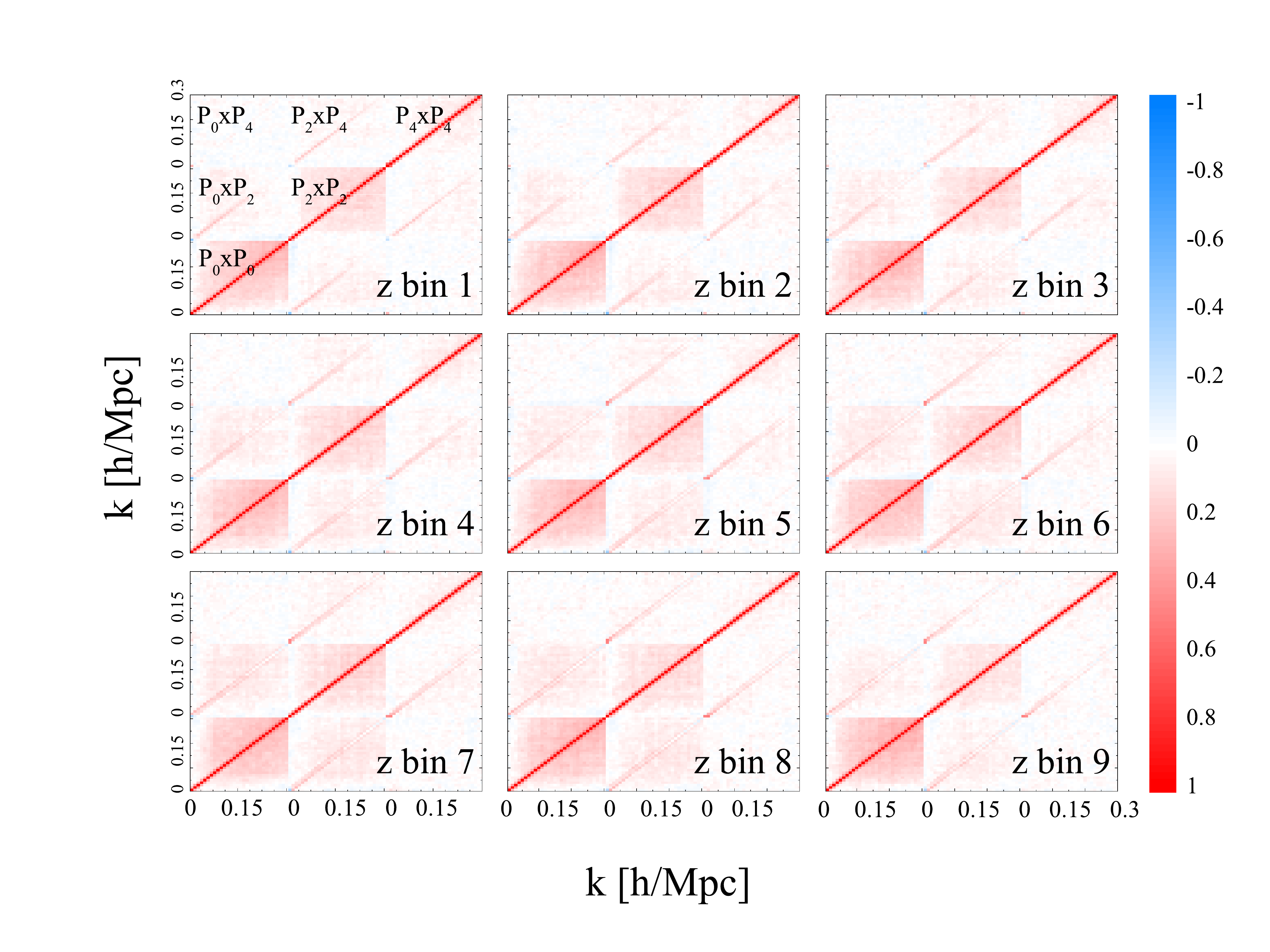}}
\caption{Same as Fig \ref{fig:corr_ngc} but for the SGC.}
\label{fig:corr_sgc}
\end{figure*}

\begin{figure*}
\centering
{\includegraphics[scale=0.5]{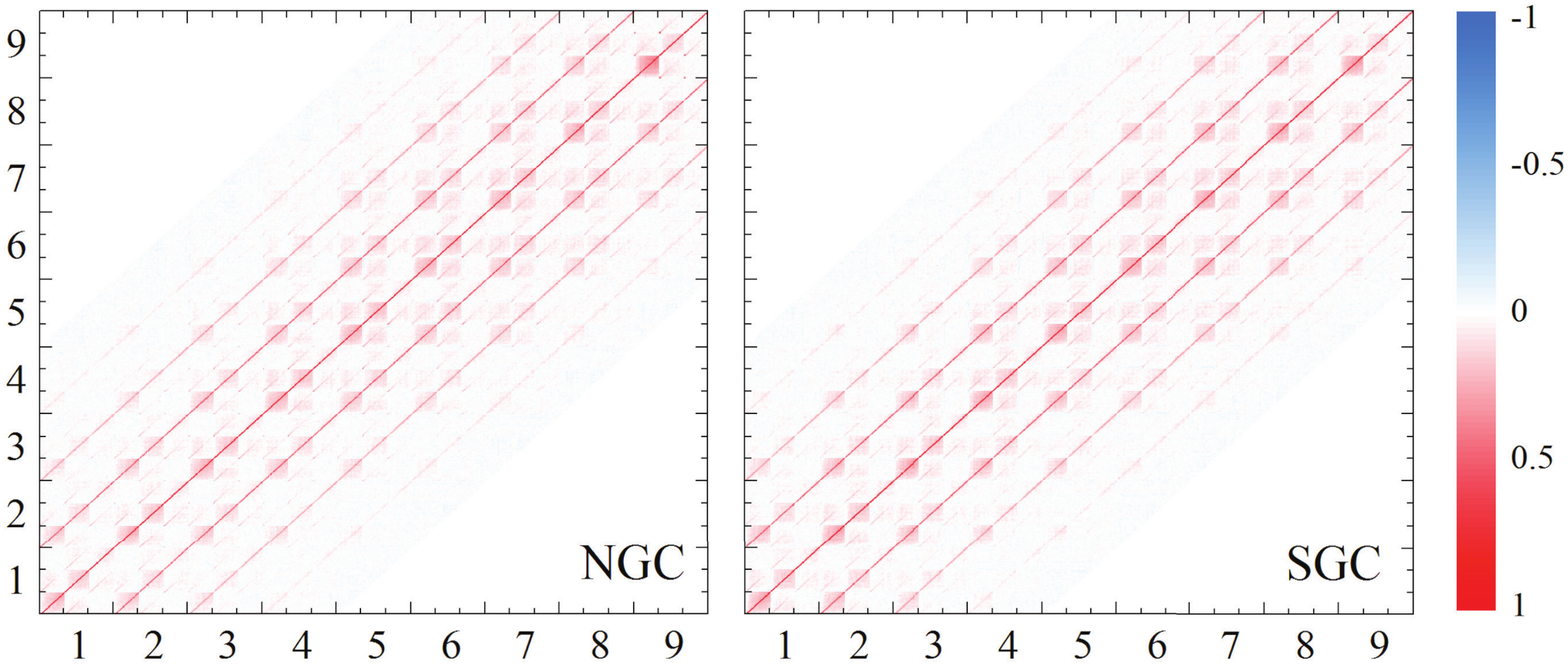}}
\caption{The full correlation matrix of $P_{\ell}(k)$ among nine redshift slices for the galaxies in the NGC (left) and in the SGC (right).}
\label{fig:corr_all}
\end{figure*}

To measure the power spectrum multipole, we need to perform Fourier transformations of the overdensity field $\Delta(\rr)$ defined in Eq (\ref{eq:Delta}) \citep{Yamamoto}. Specifically, we need to calculate the following quantity, \be\label{eq:Flk} F_{\ell}(\k)\equiv\int d\rr \ \Delta(\rr) (\hat{\k} \cdot \hat{\rr})^{\ell}  e^{i \k\cdot\rr}\ee for every $\k$ mode. This integral was recently found to be evaluable using Fast Fourier Transformations (FFTs) \citep{FFTPk, FFTPk2} instead of the expensive direct summation. We use the CAPSS package \footnote{Cosmological Analysis Package for Spectroscopic Surveys (CAPSS) is a code package developed by Gong-Bo Zhao. CAPSS is written in Fortran 90, and can be used for the measurement of galaxy power spectrum and correlation function multipoles. CAPSS is used for all the BAO analysis in this work.} which requires the FFTW library \footnote{Publicly available at \url{http://www.fftw.org/}} to perform the Fourier transformations to obtain $F_0, F_2$ and $F_4$, which is the ingredient for the measurement of the monopole, quadrupole and hexadecapole moments of the galaxy power spectrum in redshift space respectively.

Given $F_{\ell}(\k)$, the power spectrum moments can be calculated as, \footnote{The formulae presented here are an rearrangement of the original ones in \citet{FFTPk} to improve efficiency and to save memory by minimising large matrix operations.},  \be\label{eq:Pest}\hat{P}_0(k) = \int \frac{d\Omega_k}{4\pi}[F_0(\k)F_0^{\ast}(\k)]-S\ee
\be\hat{P}_2(k) = \frac{15}{2}\int \frac{d\Omega_k}{4\pi}[F_0(\k)F_2^{\ast}(\k)]-\frac{5}{2}\left[\hat{P}_0(k)+S\right]\ee
\be\hat{P}_4(k) = \frac{315}{8}\int \frac{d\Omega_k}{4\pi}[F_0(\k)F_4^{\ast}(\k)]-\frac{9}{2}\hat{P}_2(k)-\frac{63}{8}\left[\hat{P}_0(k)+S\right]\ee where $S$ is the shot noise term, which can be calculated as, 
\be\label{eq:sn} S = \sum_{i=1}^{N_{\rm G}}\left[\zeta\frac{w_{\rm T}^2}{w_{\rm fc}+w_{\rm rf}-1} + (1-\zeta)w_{\rm T}^2\right]+
\gamma^2\sum_{i=1}^{N_{\rm R}}w_{\rm FKP}^2\ee The quantity $\zeta$ is the probability that a close pair of galaxies corrected by the fibre collision weight is a true pair. We set $\zeta$ to be 0.5 following the study in \citet{fcweight}. Note that for brevity, we have dropped the dependence of all the weights on location $\rr_i$ for the $i$th galaxy or random sample in Eq (\ref{eq:sn}).

As mentioned earlier, the aliasing problem exists for all FFT-related manipulations, and it must be corrected for, especially close to the Nyquist scale. Here we follow \citet{Jing05} to correct for the aliasing effect analytically, \ie, we divide each $\k$ mode by the following correction factor for the PCS interpolation, \ba C(\k)&=&\prod_{i=1}^3 \left[1-\frac{4}{3}{\rm sin}^2\left(\frac{\pi k_i}{2 k_N}\right) + \frac{2}{5}{\rm sin}^4\left(\frac{\pi k_i}{2 k_N}\right) \right. \nn\\ && \left. - \frac{4}{315}{\rm sin}^6\left(\frac{\pi k_i}{2 k_N} \right)\right]\ea where $i$ runs over three dimensions, and $k_N$ is the Nyquist scale, $k_N = \pi N_g/L$ which is $\sim0.64\hompc$ in our case. After the anti-aliasing correction, the level of aliasing is negligible $(<0.1\%)$ on scales of interest $(k<0.3\hompc)$ of our analysis.

\subsection{The result of the $P_{\ell}(k)$ measurement}

The measurement of $P_0(k),P_2(k)$ and $P_4(k)$ for the galaxies in nine redshift slices in the NGC and SGC are shown as data points in Figs \ref{fig:pk_ngc} and \ref{fig:pk_sgc} respectively. Our measurement is in $30$ $k$ bins linearly spaced between $k=0$ to $k=0.3\hompc$. To quantify the uncertainty, we perform the same measurement on the MD-Patchy mocks, and compute the mean (shown as solid curves) and standard deviation (shown as error bars and shaded error bands) of the $P_{\ell}(k)$ measured in each $k$ bin of the 2048 mocks. We find that although the measurements in the NGC and SGC are in general consistent with each other, an offset exists. This may be due to slightly different selections used for the observations in two hemispheres. For more details of the discussion on the NGC-SGC discrepancy, we refer the readers to the companion papers of \citet{Acacia,Beutler16a,Jan16}.  

\subsection{The data covariance matrix}

The covariance between the $i$th $k$ bin of the $\ell$th order multipole in the $m$th redshift bin, and the $j$th $k$ bin of the $\ell'$th order multipole in the $n$th redshift bin can be calculated as follows, 
\ba C^{\ell,\ell'}_{ij,mn}&=&\frac{1}{N_{\rm mock}-1} \sum_{q=1}^{N_{\rm mock}} \left[P_{\ell}^q(k_i,z_m)-\bar{P}_{\ell}(k_i,z_m)\right] \times \nn \\ 
                                 && \left[P_{\ell'}^q(k_j,z_n)-\bar{P}_{\ell'}(k_j,z_n)\right], \ea where the overbars denote the average value, \ie, \be
\bar{P}_{\ell}(k_i,z_m)=\frac{1}{N_{\rm mock}} \sum_{q=1}^{N_{\rm mock}}P_{\ell}^q(k_i,z_m), \ee Here $N_{\rm mock}=2048$ is the number of mocks used. 

Note that the estimated covariance matrix using mocks needs to be corrected for a bias using the Hartlap factor \citep{Hartlap}, 
\ba \label{eq:inv_cov}
\widetilde{C}_{ij}^{-1}=f_HC_{ij}^{-1}; \ f_H=\frac{N_{\rm mock}-N_b-2}{N_{\rm mock}-1}.
\ea
where $N_b$ is the number of $k$ bins. The correction is unbiased if the error distribution of the data is Gaussian, which is only true when $N_{\rm mock} \gg N_b$ so that $f_H$ is close to 1. For covariance between $k$ bins within the same redshift slice, even if $P_0,P_2$ and $P_4$ are all included, $N_b=90$, and $f_H=0.956$. For the covariance between $k$ bins in two different redshift slices, $f_H$ reduces to $0.912$, which is also sufficiently close to $1$ \footnote{Using 30 $k$ bins for each multipole measurement is a balance between the $k$-resolution for the BAO, and the requirement that $f_H\simeq1$. This $k$-binning choice is also adopted by \citet{Acacia,Beutler16a}.}. 

Figs \ref{fig:corr_ngc} and \ref{fig:corr_sgc} show the correlation matrix (the normalised covariance matrix so that all the diagonal elements are $1$) for $P_0, P_2,P_4$ within the same redshift slice in the NGC and SGC respectively, and Fig \ref{fig:corr_all} presents the full correlation matrix among all the redshift slices. The structure of these matrices is as follows: \begin{itemize}
\item Multipoles with the same order positively correlate in general;
\item Multipoles with different orders correlate more on the same scales;
\item Multipoles in neighbouring redshift slices correlate, and the correlation generally decreases as the separation in redshift decreases;
\item Multipoles in non-overlapping redshift slices do not correlate at all.
\end{itemize}
All these observations agree with their expected behaviour: the observables correlate if they are derived using shared galaxies. 

\section{The BAO analysis}

In this section, we shall measure the isotropic and anisotropic BAO signals from the $P(k)$ multipoles and the data covariance matrix. To begin with, we describe the theoretical models, \ie, the BAO templates, for the analysis, followed by details of the fitting procedure and results.

\subsection{The template for the isotropic BAO analysis}
\label{sec:isoBAO}

\begin{figure*}
\centering
{\includegraphics[scale=0.45]{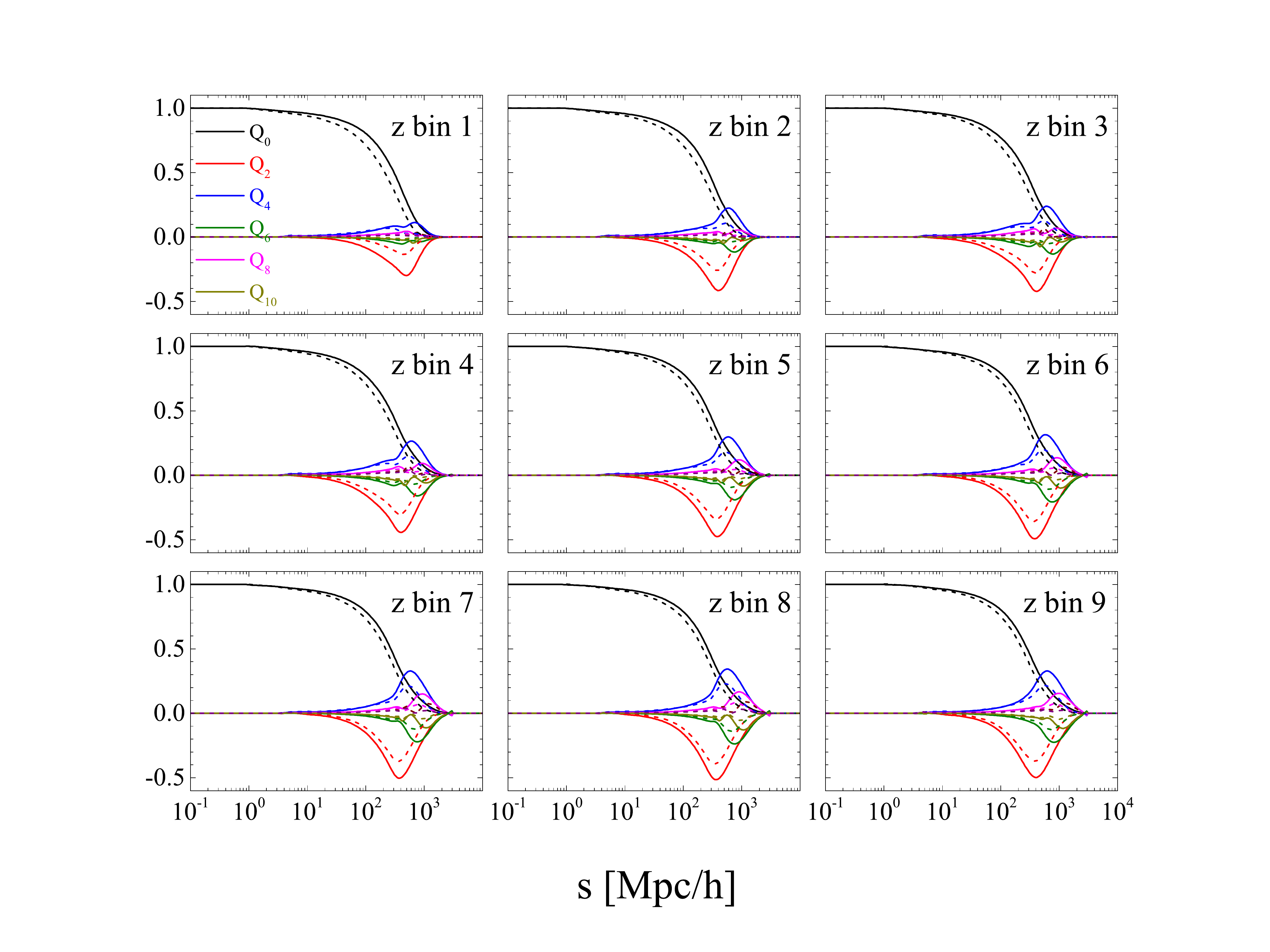}}
\caption{The normalised configuration space window function multipole $Q_{\ell}(s)$ calculated using pair counting of the random catalogues. The solid and dashed curves are for NGC and SGC respectively.}
\label{fig:window}
\end{figure*}

The isotropic BAO position can be parametrised with respect to a fiducial cosmological model using the scale dilation parameter $\alpha$,
\ba
  \alpha \equiv \frac{D_V(z)r_{d,{\rm fid}}}{D^{\rm fid}_V(z) r_d} \,,
\ea
where the volume distance $D_V(z)$ is defined in terms of the angular diameter distance $D_A(z)$ and the Hubble parameter $H(z)$, and $r_d$ is the comoving sound horizon at the drag epoch. Quantities with the super- or subscript `fid' are for the fiducial model parametrised by Eq (\ref{eq:fid}). The template for the isotropic BAO is \citep{DJE07,Beutler16a},
\be P_g(k) = P_{\rm nw} (k) \left[1+O(k) e^{-k^2 \Sigma_{\rm NL}^2/2}  \right] \ee
\be  P_{\rm nw} (k)  = B^2 P_{\rm nw, lin} (k) F(k,\Sigma_s) \ee
\be O(k) = \frac{P_{\rm lin} (k)}{P_{\rm nw, lin} (k)}-1, \ \ \ F(k)=\frac{1}{(1+k^2 \Sigma_s^2/2)} \ee where $P_{\rm lin} (k)$ is the linear power spectrum calculated using CAMB \citep{CAMB}\footnote{Available at \url{http://camb.info}}, $P_{\rm nw, lin}$ is the linear power spectrum with the BAO feature removed \citep{EH98}, $F(k,\Sigma_s)$ is the velocity damping term to account for the small scale Fingers-of-God (FoG) effect, $B$ is an overall constant for the effect of galaxy bias and redshift space distortions (RSD), $\Sigma_{\rm NL}$ quantifies the nonlinear damping scale of the oscillations. We fix $\Sigma_{\rm NL}$ to be $3.3\mpcoh$, which is motivated by numeric simulations \citep{DJE07,Seo15}. The theoretical model for the monopole is, \be\label{eq:P0} P_0(k)=\left(\frac{r_s^{\rm fid}}{r_s}\right)^3 \frac{1}{\alpha^3} P_g(k') +\frac{a_{01}}{k^3}+\frac{a_{02}}{k^2}+\frac{a_{03}}{k}+a_{04}+a_{05}k\ee where $k'=k/\alpha$. The polynomials are included here to account for systematic effects \citep{BAODR11,Acacia}. Once the parameters $\alpha, B, \Sigma_s, a_{0i}$ are known, one can use Eq (\ref{eq:P0}) to obtain a theoretical prediction for the monopole for the fitting.

\subsection{The template for the anisotropic BAO analysis}
\label{sec:anisoBAO}

The BAO feature can also be measured in both the transverse and radial directions, parametrised by $\alpha_\perp$ and $\alpha_{||}$ respectively,
\ba
\alpha_{\perp} = \frac{D_{A}(z)r_d^{\rm fid}}{D^{\rm fid}_{A}(z) r_d} \,,\,\,\,\,
\alpha_{\parallel} = \frac{H^{\rm fid}(z)r_d^{\rm fid}}{H(z) r_d} \,.
\ea
The template for the anisotropic BAO is slightly more complicated than the isotropic case due to several l.o.s.-dependent effects. The template is \citep{DJE07,Beutler16a},
\be P_g(k,\mu) = P_{\rm nw} (k,\mu) \left\{1+O(k) e^{-k^2 \left[\mu^2\Sigma_{||}^2+\left(1-\mu^2\right)\Sigma_{\perp}^2\right]/2}  \right\} \ee where $\mu$ is the cosine value of the angle between the galaxy pair separation and the l.o.s., and
\be  P_{\rm nw} (k,\mu)  = B^2 (1+\beta\mu^2)^2 P_{\rm nw, lin} (k) F(k,\mu) \ee
and \be F(k,\mu)=\frac{1}{(1+k^2\mu^2 \Sigma_s^2/2)} \ee

\ba\label{eq:AP} P_{\ell}(k)&=&\left(\frac{r_s^{\rm fid}}{r_s}\right)^3 \frac{2\ell+1}{2\alpha_{\perp}^2\alpha_{||}} \int_{-1}^{1} \d\mu \ P_g(k',\mu') \mathcal{L}_{\ell}(\mu)  \nn \\
&& +\frac{a_{\ell1}}{k^3}+\frac{a_{\ell2}}{k^2}+\frac{a_{\ell3}}{k}+a_{\ell4}+a_{\ell5}k \ea
where \ba\label{eq:kmu} && k'=\frac{k(1+\epsilon)}{\alpha} \left\{1+\mu^2 \left[\left(1+\epsilon\right)^{-6}-1\right]   \right\}^{1/2} \nn\\ 
&&\mu' = \frac{\mu}{\left(1+\epsilon\right)^3} \left\{1+\mu^2 \left[(1+\epsilon)^{-6}-1\right]   \right\}^{-1/2}\ea where \ba
\alpha = \alpha_{\perp}^{2/3} \alpha_{\parallel}^{1/3} \,, \,\,\,\,
1 + \epsilon = \left( \frac{\alpha_{\parallel}}{\alpha_{\perp}} \right)^{1/3} \,.
\ea Note that $\alpha$ is the isotropic BAO dilation, and $\epsilon$ is the warping factor. Eq (\ref{eq:AP}) shows the Alcock-Paczynski effect, which is key to allow for the simultaneous determination of $D_A$ and $H$ using multipoles of $P(k)$ \citep{BPH96}. 

The $(1+\beta\mu^2)^2$ term accounts for the RSD effect on large scales \citep{Kaiser}, and exponential damping term becomes anisotropic in this case. We set $\Sigma_\parallel=8\mpcoh$ and $\Sigma_\perp=4\mpcoh$ motivated by simulations \citep{Seo15,Beutler16a}. 

In this setup, once the parameter set ($\alpha_{\perp},\alpha_{||},B, \beta, \Sigma_s, a_{\ell i}$) is known, one can predict $P_{\ell}(k)$ using Eq (\ref{eq:AP}), thus we use Eq (\ref{eq:AP}) as a template for the anisotropic BAO fitting.

\subsection{The survey window function}

\begin{figure}
\centering
{\includegraphics[scale=0.28]{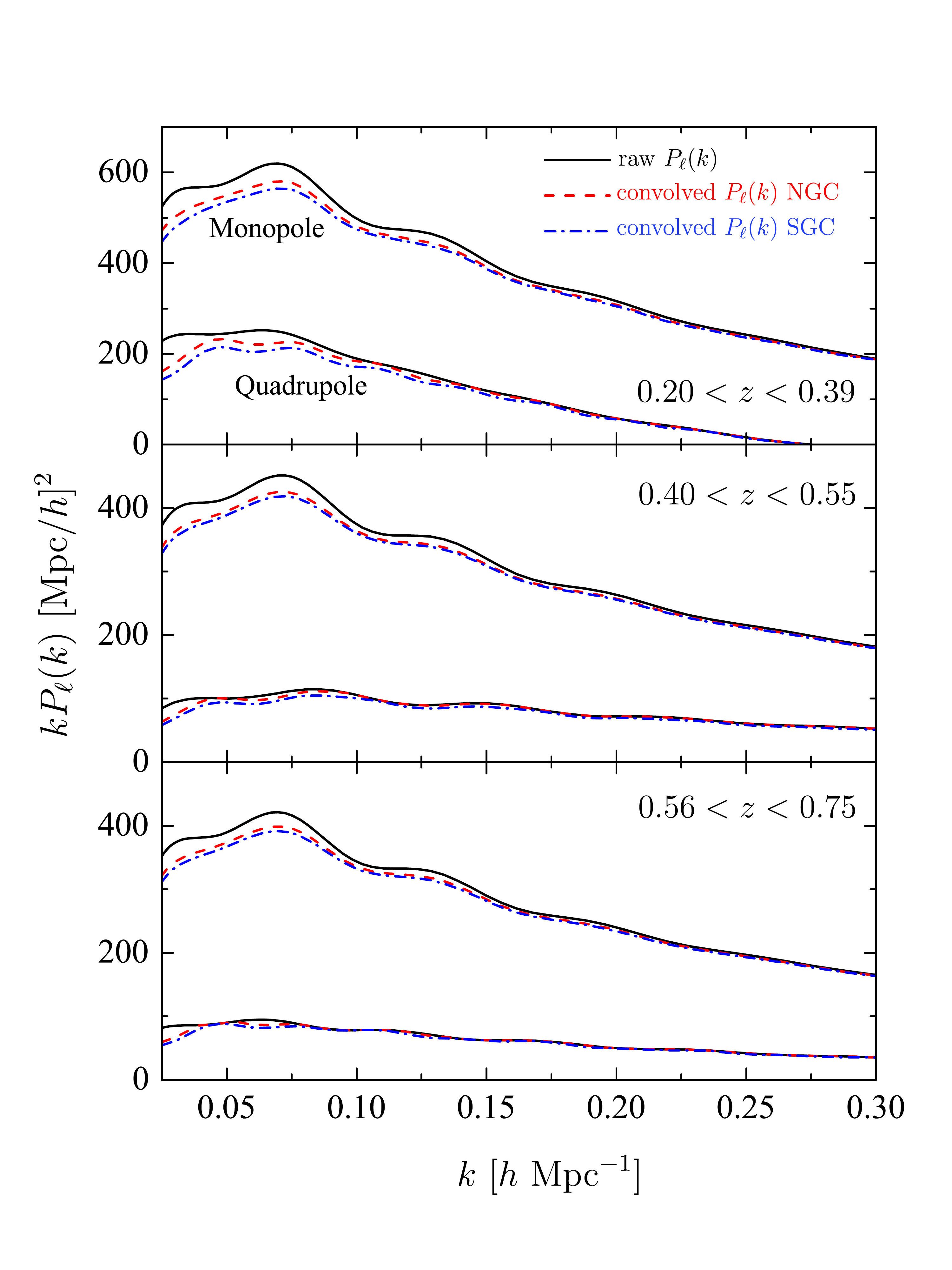}}
\caption{The theoretical $P_{\ell}(k)$ (black solid) and the convolved $P_{\ell}(k)$ with NGC (red dashed) and SGC (blue dash-dotted) window functions in three redshift slices.}
\label{fig:pk_fftlog}
\end{figure}

The theoretical model predictions derived using templates Eqs (\ref{eq:P0}) or (\ref{eq:AP}) cannot be directly compared to $P_{\ell}(k)$ measurements yet since the theoretical templates do not take into account the fact that the survey volume is irregular, and has a finite size. Ignoring these facts can overestimate the power on large scales, which may yield a biased estimate on the BAO signal. 

These effects can be accounted for by convolving the theoretical model prediction with the survey window function. The survey window function is generally anisotropic due to the irregular geometry of the survey volume, thus the window function multipoles need to be evaluated even for the isotropic BAO analysis. 

Calculating the window function multipoles and performing the three-dimensional convolution with the theoretical model prediction can be technically challenging and costly. Recently, \citet{windowpk} developed a new method for the window function evaluation and convolution. This method calculates the window function multipoles in configuration space based on a pair-counting using the random catalogue, correct for the windowing effect in real space, and then transform the result back to Fourier space using one-dimensional Hankel transformations. This is a very efficient and accurate method, thus we follow this approach in this analysis. 

We first estimate the window function multipoles from the pair counts in configuration space using a parallelised tree code in the CAPSS package,  
\be Q_{\ell}(s) \propto \int_{-1}^{1} d\mu \ RR(s,\mu)\mathcal{L}_{\ell}(\mu) \simeq \sum_{i} \ RR(s,\mu_i)\mathcal{L}_{\ell}(\mu_i) \ee where $RR$ is the pair counts of randoms at seperation $s$ with angle $\mu$, and $\mathcal{L}_{\ell}$ is the $\ell$th order Legendre polynomial. The resultant $Q_{\ell}$'s are shown in Fig \ref{fig:window}. As shown, $Q_{\ell}$ vanishes on scales $\gtrsim3000\mpcoh$, and this scale is larger in the NGC than in the SCG due to large volume in the NGC. The higher multipoles contribute less in general, which guarantees a convergence result by keeping the first few $Q_{\ell}$'s.

Given the $Q_{\ell}$'s, we compute the {\it corrected} galaxy correlation function multipoles as follows \citep{Beutler16a,Beutler16b,windowpk},
\ba\hat{\xi}_0(s)&=&\xi_0 Q_0 +\frac{1}{5}\xi_2 Q_2 +\frac{1}{9} \xi_4 Q_4 + ...\nn\\
\hat{\xi}_2(s)&=&\xi_0 Q_2 +\xi_2 \left[Q_0+\frac{2}{7}Q_2+\frac{2}{7}Q_4 \right]\nn\\
&& + \xi_4 \left[\frac{2}{7} Q_0+\frac{100}{693}Q_4+\frac{25}{143}Q_4 \right] + ...\nn\\
\hat{\xi}_4(s)&=&\xi_0 Q_4 +\xi_2 \left[\frac{18}{35} Q_2+\frac{20}{77}Q_4+\frac{45}{143}Q_6 \right]\nn\\
&& + \xi_4 \left[Q_0+\frac{20}{77} Q_2+\frac{162}{1001}Q_4+\frac{20}{143}Q_6+ \frac{490}{2431}Q_8\right] \nn\\
&&+ ...\ea
where the $\xi_{\ell}$'s are the correlation function multipoles converted from theoretical templates Eqs (\ref{eq:P0}) or (\ref{eq:AP}) using a one-dimensional Hankel transformation. Given the $\hat{\xi}_{\ell}$'s, we then perform a one-dimensional inverse Hankel transformation to obtain the window-convolved power spectrum, $P_{\ell}^{\rm conv}(k)$, using the FFTlog package \citep{fftlog} \footnote{Available at \url{http://casa.colorado.edu/~ajsh/FFTLog/}}. 

The pre- and post-convoluted monopole and quadrupole for three redshift bins are shown in Fig \ref{fig:pk_fftlog}. As expected, the window function reduces the large-scale powers due to the finite volume of the survey, and the damping effect from the SGC window function is larger due to the fact that the volume in the SGC is smaller than that in the NGC.  

\subsection{The MCMC BAO fitting}

We constrain the BAO parameters for each redshift slices using a modified version of CosmoMC \citep{cosmomc} \footnote{Available at \url{http://cosmologist.info/cosmomc/}}, which is a Markov Chain Monte Carlo (MCMC) engine. We  sample the parameter space for ${\bf p}$, which is a collection of BAO parameters explained in Sections \ref{sec:isoBAO} and \ref{sec:anisoBAO}, by minimising the following $\chi^2$,
\ba
 \chi^2 (\bold{p}) \equiv  \sum_{i,j}^{\ell,\ell'}  \left[P^{\rm conv}_{\ell} (k_i, \bold{p}) -P_{\ell}(k_i) \right] F^{\ell,\ell'}_{ij} \left[P_{\ell'}^{\rm conv}(k_j, \bold{p}) -P_{\ell'}(k_j)\right] \nonumber
\ea
where $F^{\ell,\ell'}_{ij}$ is the inverse of the data covariance matrix. Note that when using both the NGC and SGC data for the constraint, we use two separate $B$ parameters for the NGC and SGC to account for the offset discussed earlier. We analytically marginalise over the coefficients of polynomials in each MCMC step, \ie, we calculate the optimal values of the coefficients given a set of parameters to minimise the $\chi^2$, \be\label{eq:chi2_c}\chi^2=({\bf D}+{\bf X})^T {\bf F}({\bf D}+{\bf X})\ee where the residue vector ${\bf D}$ is defined as
\be D(k) \equiv P^{\rm data}(k) - P^{\rm theo.}(k)\ee and the polynomial vector $X$ is,
\be {\bf X} \equiv  {\bf A} \cdot {\bf K}  \ee where \be{\bf K}\equiv\left(\frac{1}{k^3},\frac{1}{k^2},\frac{1}{k},{1},k \right)\ee Given ${\bf D}$ and ${\bf F}$ at each MCMC step, our aim is to analytically determine the coefficients vector ${\bf A}$ to minimise $\chi^2$. To do this, we expand Eq (\ref{eq:chi2_c}), and setting $\partial\chi^2/\partial {\bf X}=0$ yields,
\be {\bf A}^T = ({\bf K F K}^T)^{-1} {\bf K F D}\ee This procedure can avoid fitting these weakly constrained nuisance parameters, making the MCMC chains much easier to converge.  

After the MCMC chains converge, we perform statistics on the chain elements to obtain the posterior distribution of each parameter, and the correlation among parameters. Note that the data covariance matrix estimated from the finite mocks inevitably has errors, which propagate into errors of the parameters. To correct for this, we follow \citet{Percival} and rescale the variance of each parameter by 
\ba
M=\sqrt{\frac{1+B(N_b-N_p)}{1+A+B(N_p+1)}}
\ea
where $N_p$ and $N_b$ are the number of parameters and number of $k$ bins respectively, and 
\ba
&&A=\frac{2}{(N_{\rm mock}-N_b-1)(N-N_b-4)}, \nn \\
&&B=\frac{N_{\rm mock}-N_b-2}{(N_{\rm mock}-N_b-1)(N_{\rm mock}-N_b-4)}.
\ea

\subsection{Mock tests}
\label{sec:mock}

\begin{figure}
\centering
{\includegraphics[scale=0.33]{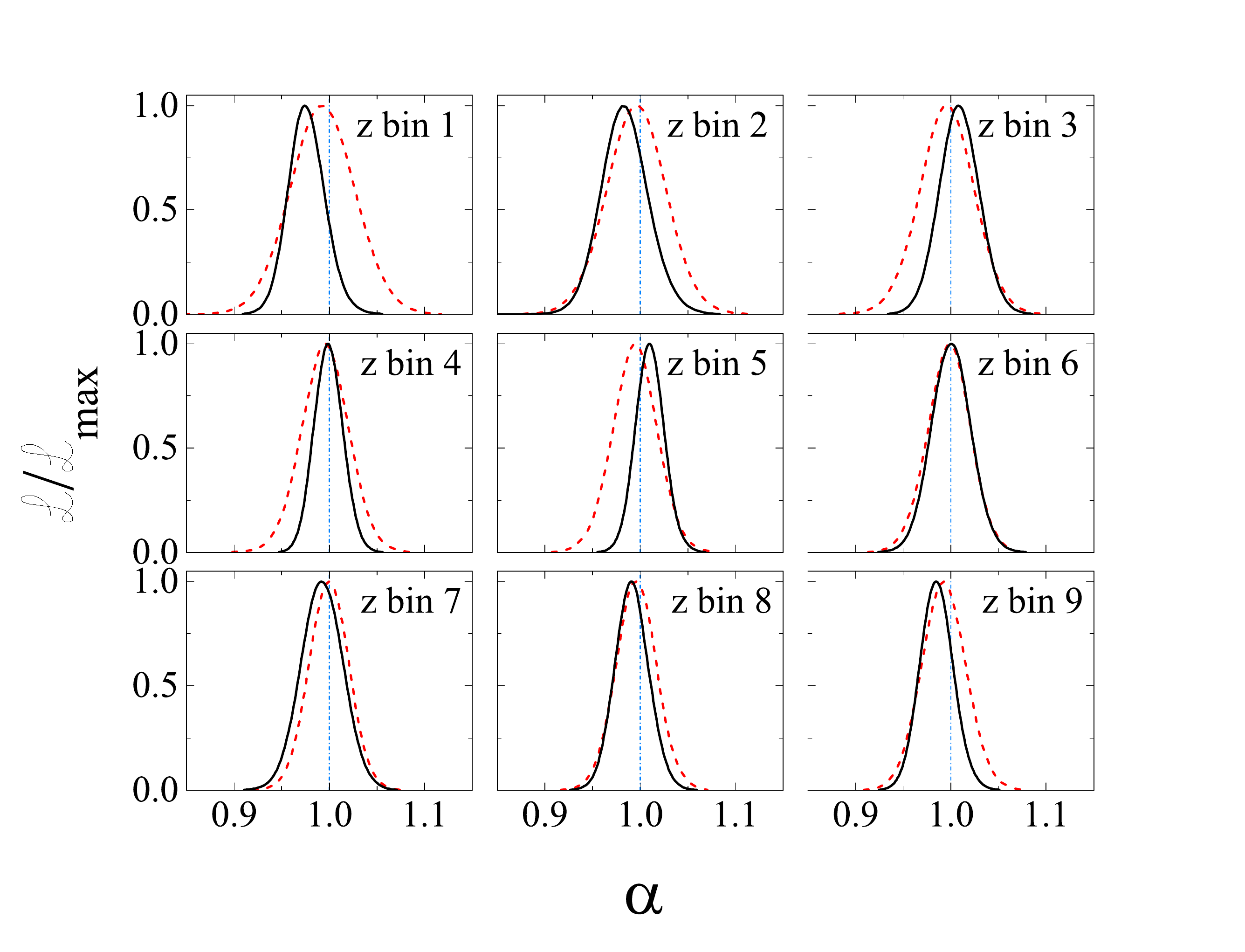}}
\caption{The one-dimensional posterior distribution of the isotropic $\alpha$ derived from the observations (black solid) and mock catalogues (red dashed) respectively. The blue dash-dotted lines show $\alpha=1$ for a reference. }
\label{fig:alpha}
\end{figure}

\begin{figure*}
\centering
{\includegraphics[scale=0.62]{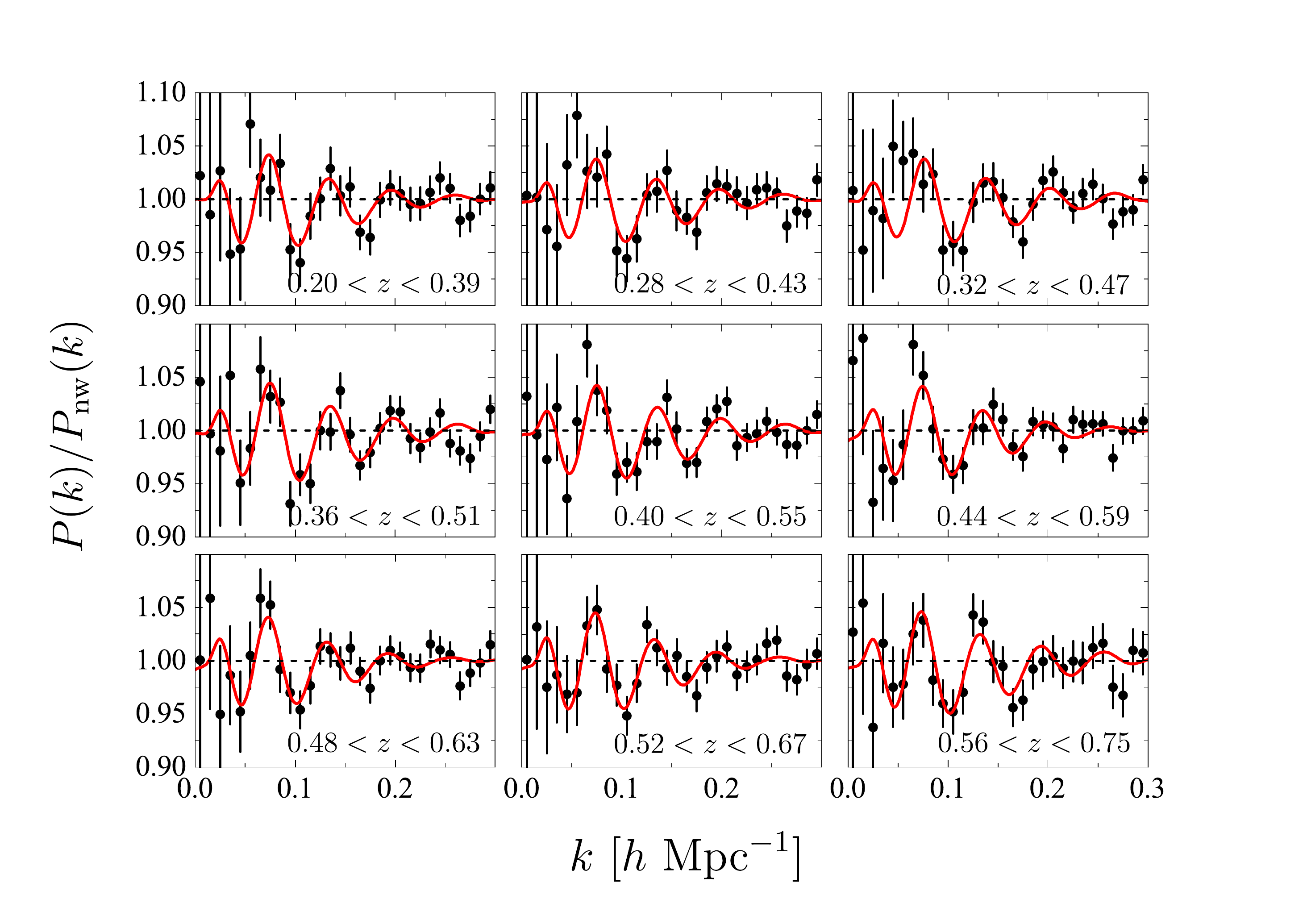}}
\caption{An overplot of the measured $P(k)$ monopole using the galaxies in the NGC (data with error bars) and the best fit model (red solid), rescaled by the best fit model without the BAO feature. }
\label{fig:bao_ratio}
\end{figure*}

We first validate our pipeline by performing the BAO analysis on the MD-Patchy mocks. We fit the isotropic, and anisotropic BAO parameters to the average of 2048 mocks \footnote{We fit the mean of 2048 mocks in the same way as we fit the observational data.}. The isotropic BAO test is shown in Table \ref{tab:isobao} and Fig \ref{fig:alpha} (red dashed curves for one-dimensional posterior distribution of $\alpha$). As shown in the left part of Table \ref{tab:isobao}, the mean value of $\alpha$'s are consistent with 1, which is the input value of all the mocks, within $0.3\sigma$ in the worst case (for redshift bin $z_9$). The shift from 1 could be due to nonlinearities such as the mode-coupling effect on quasi-nonlinear scales \citep{modecoupling}, which is not included in our fitting templates, but can be approximately estimated analytically \citep{Seo08,modecoupling}. The expected shift in $\alpha$ in 9 redshift bins are shown in the $\Delta\alpha_{\rm MC}$ column in Table \ref{tab:isobao}, which is $0.17\sigma$ in the worst case (for redshift bin $z_7$). To account for this systematic effect conservatively, we include a systematic error budget on $\alpha$ by adding $\Delta\alpha_{\rm MC}$ and the shift of the mean $\alpha$ from 1 in quadrature, for the BAO measurements using the galaxy sample, which will be presented later.  

\begin{figure*}
\centering
{\includegraphics[scale=0.4]{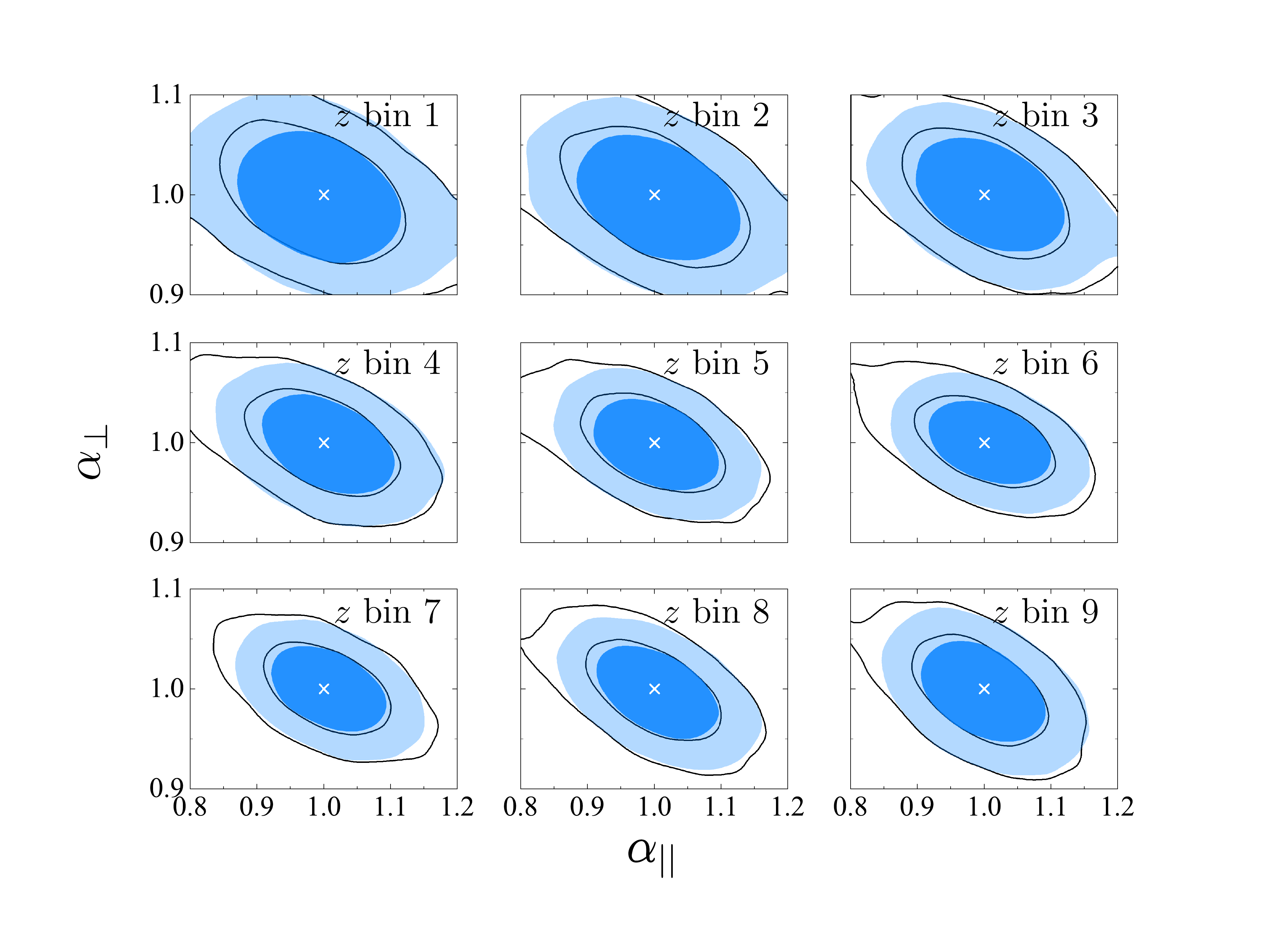}}
\caption{The 68 and 95\% CL contour plots for $\alpha_{\parallel}$ and $\alpha_{\perp}$ using $P(k)$ multipoles (black unfilled contours: $P_0+P_2$; blue filled contours: $P_0+P_2+P_4$) measured from the MD-PATCHY mock catalogue in nine redshift slices. The unfilled black and filled blue contours are results using galaxies in the NGC, and all galaxies in the catalogue respectively. The white cross in each panel illustrates the fiducial model ($\alpha_{\parallel}=\alpha_{\perp}=1$).}
\label{fig:bao2d_mock}
\end{figure*}

The mock test results for anisotropic BAO are shown in Table \ref{tab:2dbaomock} and in Fig \ref{fig:bao2d_mock}. As shown, the hexadecapole improves the constraint in all redshift bins, \ie, 
\begin{itemize}
\item It shrinks the statistical error budget by $9-15\%$ for $\alpha_{||}$, and $8-11\%$ for $\alpha_{\perp}$;
\item It generally makes the mean value of both $\alpha_{||}$ and $\alpha_{\perp}$ more consistent with unity;
\item It reduces the degeneracy between $\alpha_{||}$ and $\alpha_{\perp}$ by $18-32\%$.
\end{itemize}

This means that the hexadecapole from DR12 sample is indeed informative for BAO studies. It is true that given the level of uncertainty of $P_4(k)$, the BAO feature is barely visible. However, it can improve the global fitting by providing constraints on the amplitude parameters $B$ and the RSD parameter $\beta$, and thus reduce the degeneracy between $\alpha_{||}$ and $\alpha_{\perp}$, and improve their constraints indirectly. Given the importance of the hexadecapole, we shall include it in all the analysis in this work unless otherwise mentioned. 

We quantify the systematic error budget similarly to the isotropic BAO case, \ie, the systematic error is estimated using the quadrature addition between the bias caused by the mode-coupling effect, and the shift of mean $\alpha_{||}$ and $\alpha_{\perp}$ from unity. The mode-coupling bias is taken to be $\Delta\alpha_{||}^{\rm MC}=0.001$ and $\Delta\alpha_{\perp}^{\rm MC}=0.009$ \citep{Ross16}. This yields a $0.15-0.76\%$ systematic error on $\alpha_{||}$, and $0.09-0.1\%$ on $\alpha_{\perp}$.

In summary, we validate our pipeline using mock tests, namely, the bias introduced by the pipeline is small compared to the statistical error in all cases, and the bias is accounted for by the systematic error budget. 

\begin{table*}
\caption{The constraint on the isotropic BAO parameters $\alpha$ and $D_V$, from the mocks (left) and the galaxy catalogues (right). For quantities with double error bars, the first and second shows the statistical and systematical error budget respectively.}
\begin{center} 
\begin{tabular}{cccccc}
\hline\hline
                    & \multicolumn{2} {c} {Mock catalogue} &   \multicolumn{3} {c} {Galaxy catalogue}  \\ 
                    \hline 
$z$  bins   &   $\Delta\alpha^{\rm MC}$   &$ \alpha $    & $ \alpha $ &  $\left(\frac{D_V}{\rm Mpc}\right)\left(\frac{r_d^{\rm fid}}{r_d}\right)$   & $\chi^2/{\rm dof}$\\ \hline
$z_1$	&$	0.0040	$&$	0.9925	\pm	0.0346	$&$	0.9765	\pm	0.0190	\pm	0.0085	$&$	1208.36	\pm	23.51	\pm	10.30	$	&	$	43/51	$	\\
$z_2$	&$	0.0039	$&$	0.9960	\pm	0.0308	$&$	0.9822	\pm	0.0332	\pm	0.0056	$&$	1388.36	\pm	46.93	\pm	7.790  	$	&	$	59/51	$	\\
$z_3$	&$	0.0039	$&$	0.9956	\pm	0.0283	$&$	1.0088	\pm	0.0205	\pm	0.0058	$&$	1560.06	\pm	31.70	\pm	9.120  	$	&	$	67/51	$	\\
$z_4$	&$	0.0038	$&$	0.9955	\pm	0.0245	$&$	0.9992	\pm	0.0149	\pm	0.0059	$&$	1679.88	\pm	25.05	\pm	9.850  	$	&	$	69/51	$	\\
$z_5$	&$	0.0037	$&$	0.9945	\pm	0.0231	$&$	1.0102	\pm	0.0149	\pm	0.0066	$&$	1820.44	\pm	26.85	\pm	12.04 	$	&	$	69/51	$	\\
$z_6$	&$	0.0036	$&$	0.9979	\pm	0.0221	$&$	1.0003	\pm	0.0204	\pm	0.0041	$&$	1913.54	\pm	39.03	\pm	7.930  	$	&	$	50/51	$	\\
$z_7$	&$	0.0035	$&$	0.9994	\pm	0.0206	$&$	0.9923	\pm	0.0216	\pm	0.0035	$&$	2001.91	\pm	43.58	\pm	7.050  	$	&	$	56/51	$	\\
$z_8$	&$	0.0034	$&$	0.9958	\pm	0.0209	$&$	0.9914	\pm	0.0175	\pm	0.0054	$&$	2100.43	\pm	37.08	\pm	11.29	$	&	$	51/51	$	\\
$z_9$	&$	0.0032	$&$	0.9926	\pm	0.0229	$&$	0.9852	\pm	0.0171	\pm	0.0081	$&$	2207.51	\pm	38.32	\pm	17.80	$	&	$	50/51	$	\\ \hline\hline           
\end{tabular}
\end{center}
\label{tab:isobao}
\end{table*}

\begin{table*}
\caption{The constraint on the anisotropic BAO signal, $\alpha_{\perp}$  and $\alpha_{\parallel}$, and their correlation coefficient, $r_{\perp \parallel}$.}
\begin{center} 
\begin{tabular}{ccccccc}
\hline\hline 
        & &   Mock catalogue  ($P_0+P_2$)         &                    &    & Mock catalogue  ($P_0+P_2+P_4$) &\\ \hline
$z$ bins & $\alpha_{\parallel}$ &  $\alpha_{\perp}$  & $r_{\parallel \perp }$ & $\alpha_{\parallel}$ &$ \alpha_{\perp}$ & $r_{\parallel \perp}$  \\ \hline
$z_1$	&	$	0.9841	\pm	0.0855	$	&	$	1.0023	\pm	0.0459	$	&	$	-0.44	        $	&	$	0.9928	\pm	0.0768	$	&	$	0.9970	\pm	0.0416	$	&	$	-0.30 	$	\\
$z_2$	&	$	0.9985	\pm	0.0861	$	&	$	0.9990	\pm	0.0449	$	&	$	-0.49 	$	&	$	1.0046	\pm	0.0763	$	&	$	0.9952	\pm	0.0405	$	&	$	-0.35 	$	\\
$z_3$	&	$	1.0008	\pm	0.0796	$	&	$	1.0024	\pm	0.0410	$	&	$	-0.49 	$	&	$	1.0072	\pm	0.0705	$	&	$	0.9991	\pm	0.0370	$	&	$	-0.37 	$	\\
$z_4$	&	$	0.9942	\pm	0.0735	$	&	$	1.0010	\pm	0.0344	$	&	$	-0.49 	$	&	$	1.0047	\pm	0.0641	$	&	$	0.9976	\pm	0.0317	$	&	$	-0.38 	$	\\
$z_5$	&	$	0.9948	\pm	0.0702	$	&	$	1.0001	\pm	0.0324	$	&	$	-0.50 	$	&	$	1.0020	\pm	0.0598	$	&	$	0.9977	\pm	0.0295	$	&	$	-0.38 	$	\\
$z_6$	&	$	0.9972	\pm	0.0683	$	&	$	1.0021	\pm	0.0303	$	&	$	-0.51 	$	&	$	1.0069	\pm	0.0581	$	&	$	0.9996	\pm	0.0269	$	&	$	-0.36 	$	\\
$z_7$	&	$	1.0034	\pm	0.0628	$	&	$	1.0008	\pm	0.0296	$	&	$	-0.50 	$	&	$	1.0075	\pm	0.0548	$	&	$	0.9996	\pm	0.0274	$	&	$	-0.39 	$	\\
$z_8$	&	$	0.9971	\pm	0.0659	$	&	$	0.9990	\pm	0.0329	$	&	$	-0.55 	$	&	$	1.0049	\pm	0.0582	$	&	$	0.9962	\pm	0.0296	$	&	$	-0.45 	$	\\
$z_9$	&	$	0.9913	\pm	0.0654	$	&	$	0.9994	\pm	0.0354	$	&	$	-0.51 	$	&	$	0.9989	\pm	0.0598	$	&	$	0.9965	\pm	0.0324	$	&	$	-0.40 	$	\\

\hline\hline
\end{tabular}
\end{center}
\label{tab:2dbaomock}
\end{table*}

\begin{table*}
\caption{The mean value with 68\% statistical error (first error bar) and systematic error (second error bar) of the anisotropic BAO signal, $\alpha_{\perp}$, $\alpha_{\parallel}$, $Hr_d$ and $D_A/r_d$, the corresponding correlation coefficient, and the reduced $\chi^2$ to quantify the goodness-of-fit.}
\begin{center} 
\begin{tabular}{ccccccc}
\hline\hline 
$z$ bins &  $\alpha_{\parallel}$ &$ \alpha_{\perp}$ & $r_{\parallel \perp}$   & $\left(\frac{H}{\rm km \ s^{-1}  \ Mpc^{-1}}\right)\left(\frac{r_d}{r_d^{\rm fid}}\right)$ &  $\left(\frac{D_A}{\rm Mpc}\right)\left(\frac{r_d^{\rm fid}}{r_d}\right)$  & $\chi^2/{\rm dof}$ \\ \hline
$z_1$      & 	$	1.0214	\pm	0.0522	\pm	0.0073	$ &$	0.9592	\pm	0.0402	\pm	0.0095	$&$		-0.43		$&$		78.30	\pm	4.07		\pm	0.57		$&$	931.420	\pm	39.42	\pm	8.840	$ &  $150/144$\\
$z_2$      & 	$	1.0687	\pm	0.0694	\pm	0.0047	$ &$	0.9751	\pm	0.0322	\pm	0.0102	$&$		-0.23		$&$		77.20	\pm	5.30		\pm	0.36		$&$	1047.04	\pm	33.65	\pm	10.68	$ & $156/144$\\
$z_3$      & 	$	1.0583	\pm	0.0539	\pm	0.0073	$ &$	0.9878	\pm	0.0280	\pm	0.0090	$&$		-0.35		$&$		79.72	\pm	4.27		\pm	0.58		$&$	1131.34	\pm	34.06	\pm	10.23	$& $180/144$\\
$z_4$      & 	$	1.0751	\pm	0.0396	\pm	0.0048	$ &$	0.9785	\pm	0.0172	\pm	0.0093	$&$		-0.30		$&$		80.29	\pm	2.96		\pm	0.39		$&$	1188.78	\pm	20.90	\pm	11.07	$&$184/144$\\
$z_5$      & 	$	1.0432	\pm	0.0389	\pm	0.0022	$ &$	0.9985	\pm	0.0189	\pm	0.0093	$&$		-0.25		$&$		84.69	\pm	3.21		\pm	0.19		$&$	1271.43	\pm	24.03	\pm	11.81	$& $173/144$\\
$z_6$      &	$	0.9865	\pm	0.0743	\pm	0.0070	$ &$	1.0093	\pm	0.0202	\pm	0.0090	$&$		-0.37		$&$		91.97	\pm	6.85		\pm	0.64		$&$	1336.53	\pm	26.72	\pm	12.04	$& $149/144$\\
$z_7$      &	$	0.9526	\pm	0.0710	\pm	0.0076	$ &$	1.0116	\pm	0.0205	\pm	0.0090	$&$		-0.26		$&$		97.30	\pm	7.16		\pm	0.74		$&$	1385.47	\pm	28.04	\pm	12.48	$& $165/144$\\
$z_8$      &	$	0.9735	\pm	0.0528	\pm	0.0050	$ &$	1.0085	\pm	0.0217	\pm	0.0098	$&$		-0.35		$&$		97.07	\pm	5.24	 	\pm	0.49		$&$	1423.43	\pm	30.66	\pm	13.91	$& $144/144$\\
$z_9$      &	$	0.9931	\pm	0.0474	\pm	0.0015	$ &$	0.9932	\pm	0.0378	\pm	0.0097	$&$		-0.56		$&$		97.70	\pm	4.58		\pm	0.15		$&$	1448.81	\pm	55.12	\pm	13.99	$& $138/144$\\  
\hline  \hline              
\end{tabular}
\end{center}
\label{tab:anisobao}
\end{table*}

\subsection{BAO measurements from DR12 sample}

In this section, we shall apply our BAO analysis pipeline on the DR12 sample, and present the main results of this paper. 

\subsubsection{Isotropic BAO measurements}

The isotropic BAO fitting result is shown in the right part of Table \ref{tab:isobao} and in Figs \ref{fig:alpha} and \ref{fig:bao_ratio} (black solid). Fig \ref{fig:bao_ratio} displays the best-fit monopole and data points, divided by the smoothed power spectrum. As shown, the BAO signal is well extracted in all the redshift slices. From Table \ref{tab:isobao} and Fig \ref{fig:alpha}, which shows the one-dimensional posterior distribution of $\alpha$, in comparison to those measured from the mocks, we see that the isotropic BAO distance is determined at a precision of $1.5\%$ to $3.4\%$, depending on the effective redshifts. We also notice that $\alpha$ in three redshift slices deviate from 1 at $\gtrsim 1\sigma$ level. This may suggest that the fiducial cosmology, which is the $\Lambda$CDM model with parameters listed in Eq (\ref{eq:fid}), might be in tension with the DR12 galaxy sample. We shall explore this more in a companion paper \citep{Zhao16}. 

\subsubsection{Anisotropic BAO measurements}

The anisotropic BAO measurements are presented in Tables \ref{tab:anisobao}, Figs \ref{fig:baoring} to \ref{fig:DM_H}. Table \ref{tab:anisobao} shows the constraint on $\alpha_{||}$ and $\alpha_{\perp}$, $D_A/r_d$ and $H \ r_d$ at nine effective redshifts with the correlation coefficients and the reduced $\chi^2$ to quantify the goodness-of-fit. We can see that the anisotropic BAO distances in terms of $D_A/r_d$ and $H \ r_d$ are measured to a precision of $1.8\% -4.2\%$ and 
$3.7\% - 7.5\%$ respectively, depending on the effective redshifts. The reduced $\chi^2$ is sufficiently close to unity in all cases, which means that the fitting result is as expected. We also notice that the $\alpha$'s show deviation from 1 at $\gtrsim 1\sigma$ level, which is consistent with the result of the isotropic BAO measurement. 

\begin{figure*}
\centering
{\includegraphics[scale=0.4]{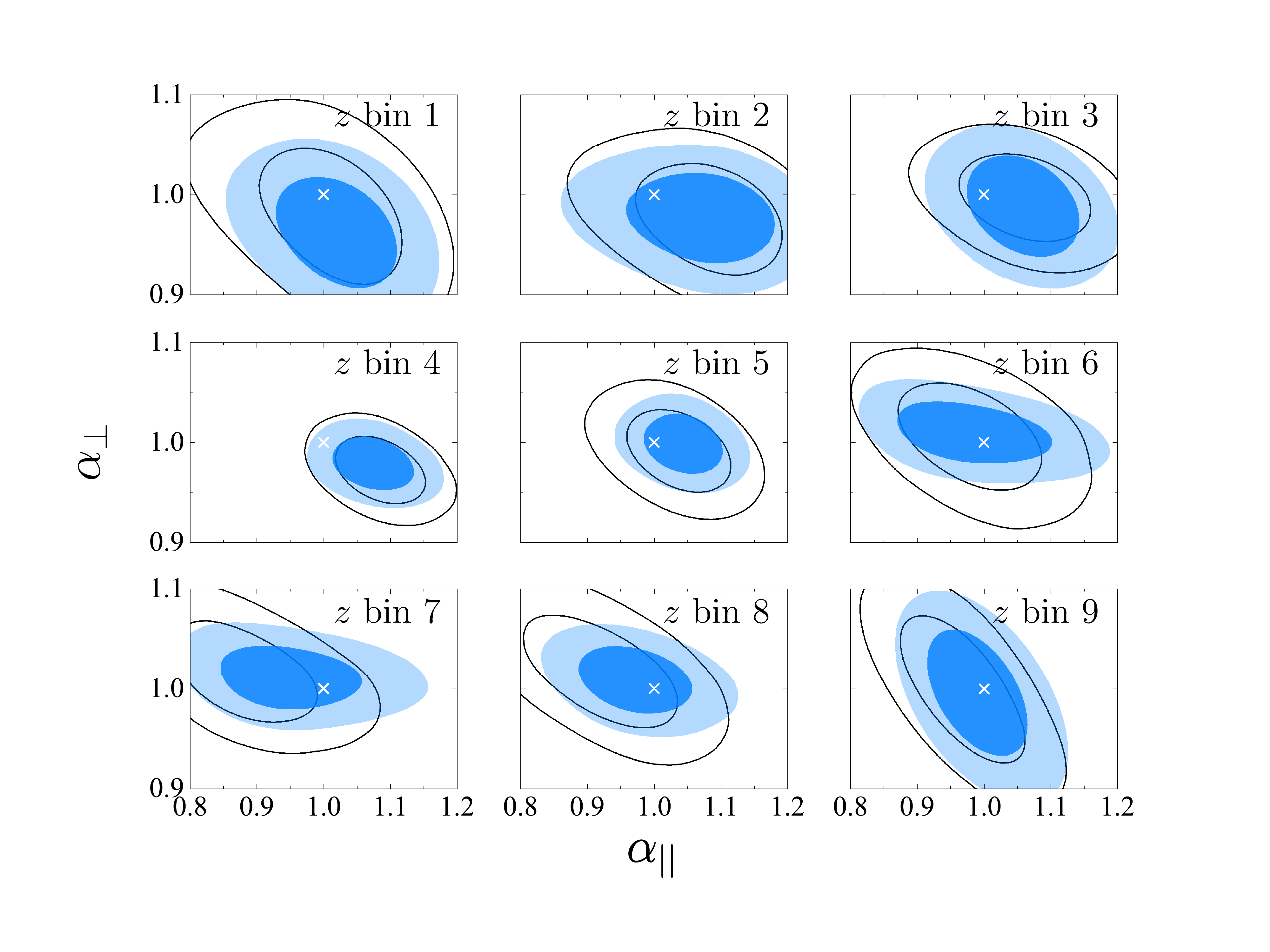}}
\caption{The 68 and 95\% CL contour plots for $\alpha_{\parallel}$ and $\alpha_{\perp}$ using $P(k)$ multipoles measured from the DR12 galaxy sample in nine redshift slices. The unfilled black and filled blue contours are results using galaxies in the NGC, and all galaxies in the catalogue respectively. The white cross in each panel illustrates the fiducial model ($\alpha_{\parallel}=\alpha_{\perp}=1$).}
\label{fig:bao2d_data}
\end{figure*}

Fig \ref{fig:bao2d_data} shows the contour plots between $\alpha_{||}$ and $\alpha_{\perp}$ using galaxies in the NGC (unfilled black) and NGC+SGC (filled blue). These results show that the BAO distances measured from the NGC and SGC are in general consistent with each other, and complementary. In the two-dimensional plane, we see the deviation from the fiducial model (shown as white crosses) at $\gtrsim 1\sigma$ level only in the fourth redshift slice. 

\begin{figure*}
\centering
{\includegraphics[scale=0.3]{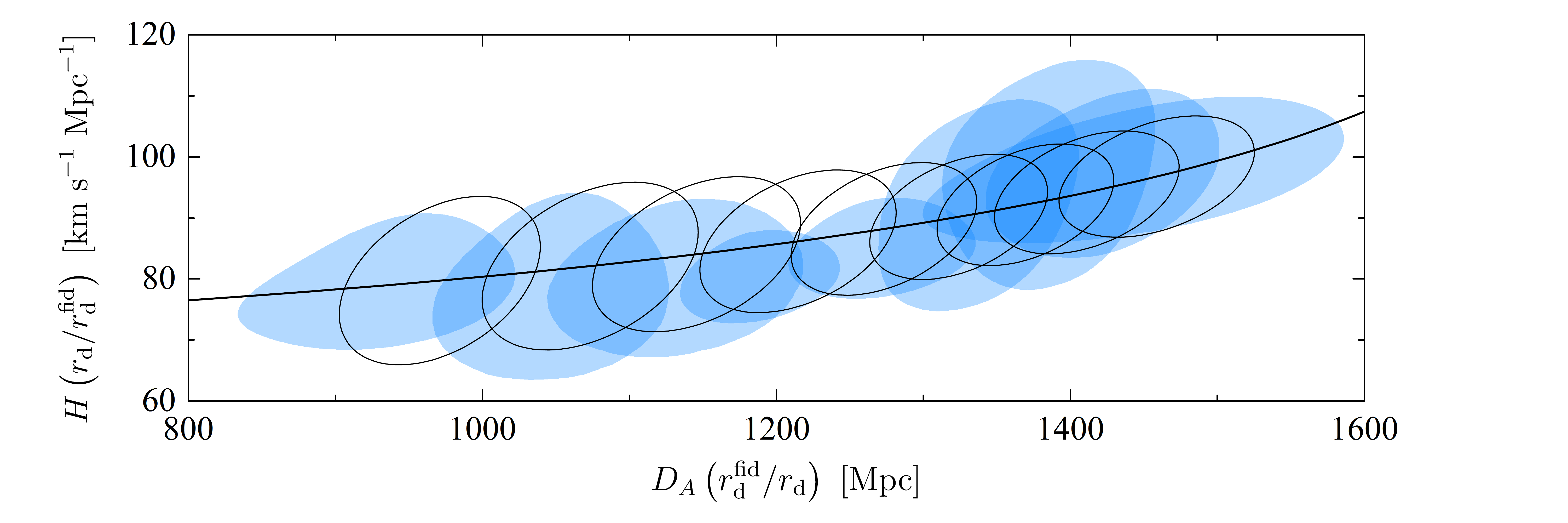}}
\caption{Shaded contours: the 95\% CL contour plots for $D_A(r_{\rm d}^{\rm fid}/r_{\rm d})$ and $H(r_{\rm d}/r_{\rm d}^{\rm fid})$ derived from DR12 galaxies in nine redshift slices; black unfilled contours: the Fisher matrix forecast. For contours from left to right, the effective redshifts of galaxies used increase from $z_{\rm eff}=0.31$ to $z_{\rm eff}=0.64$. The black solid curve shows the prediction of the fiducial model used in this analysis.}
\label{fig:DA_H_gal}
\end{figure*}

\begin{figure}
\centering
{\includegraphics[scale=0.2]{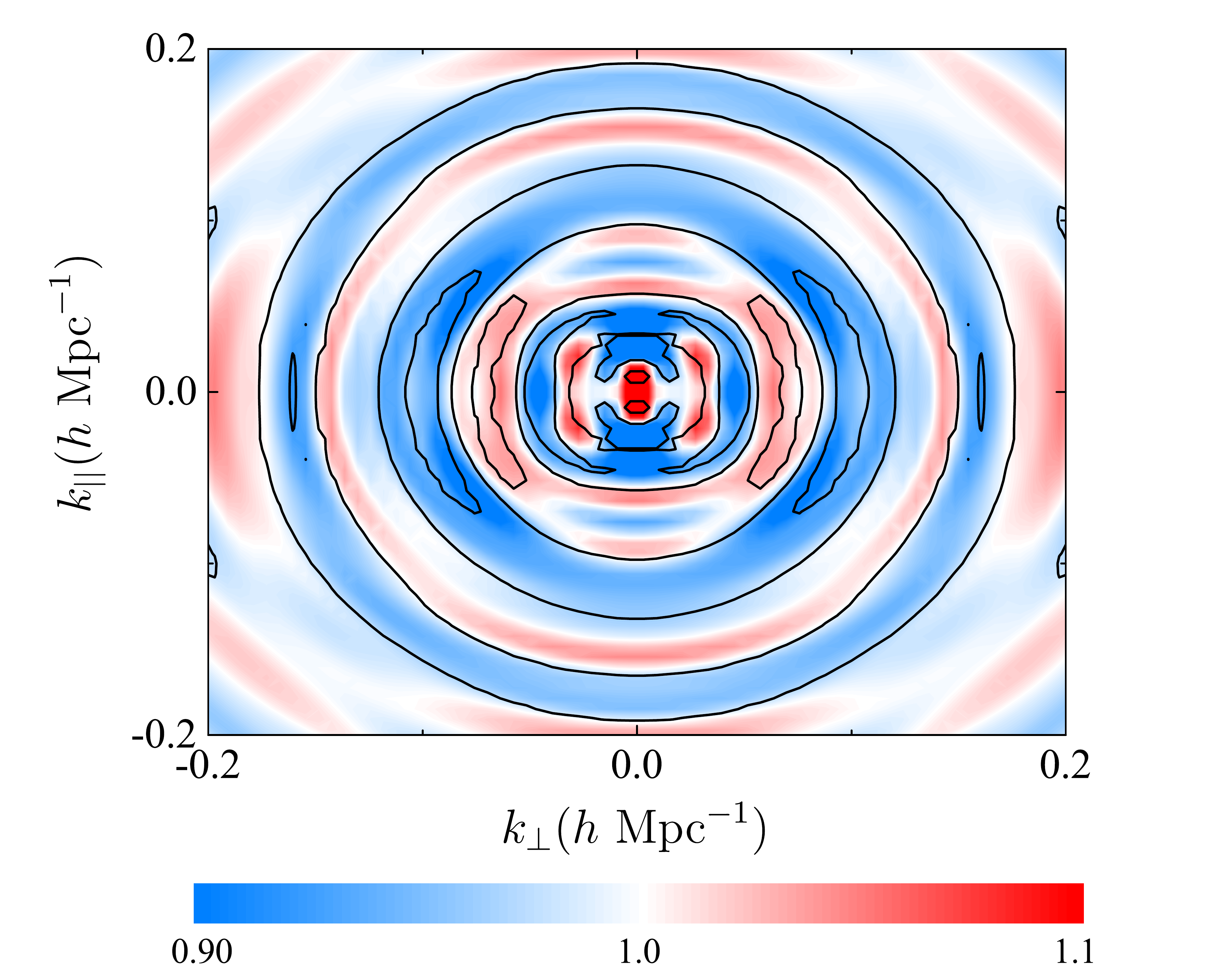}}
\caption{The 68 and 95\% CL contour plots for $\alpha_{\parallel}$ and $\alpha_{\perp}$ using $P(k)$ multipoles measured from the DR12 galaxy sample in nine redshift slices. The unfilled black and filled blue contours are results using galaxies in the NGC, and all galaxies in the catalogue respectively. The white cross in each panel illustrates the fiducial model ($\alpha_{\parallel}=\alpha_{\perp}=1$).}
\label{fig:baoring}
\end{figure}

\begin{figure}
\centering
{\includegraphics[scale=0.22]{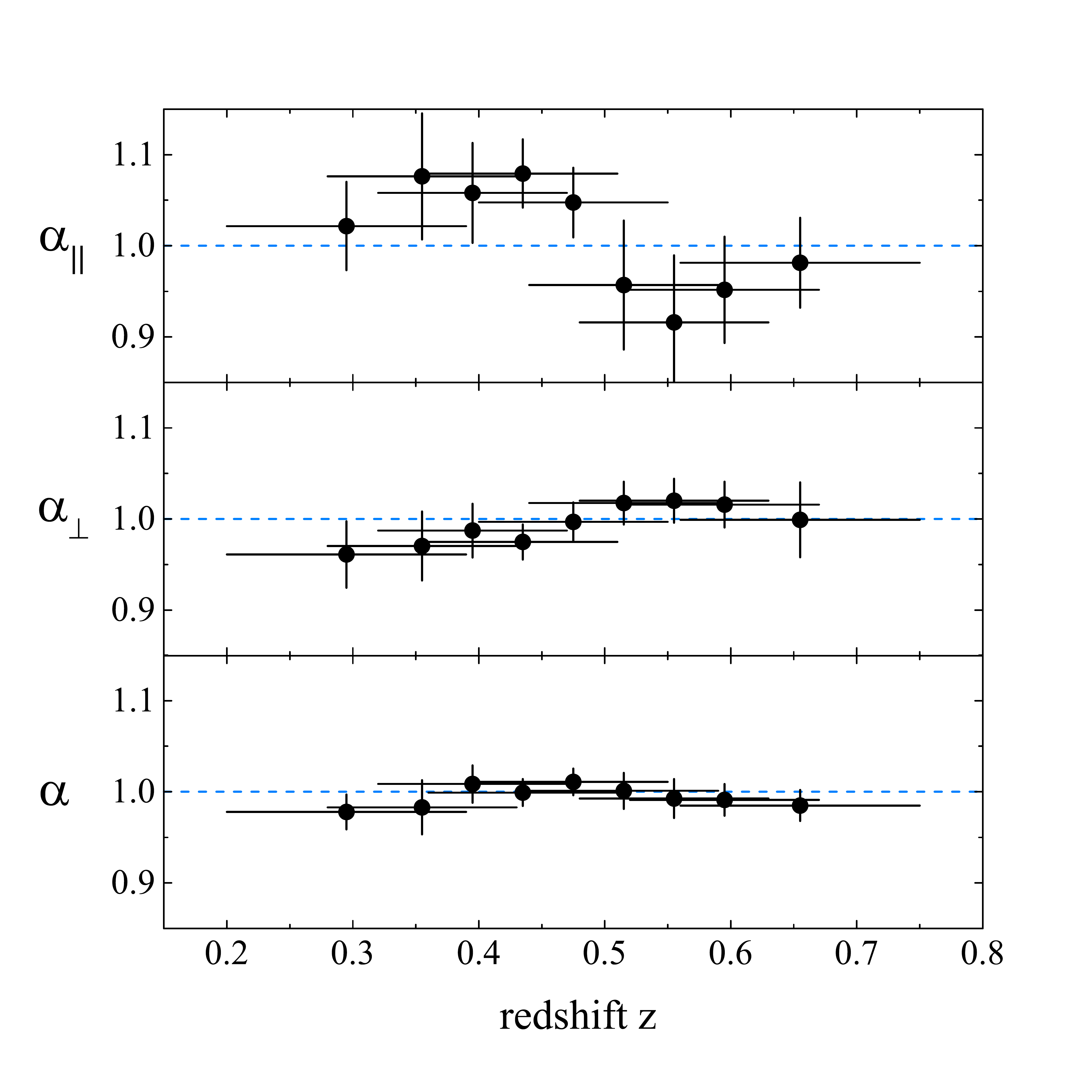}}
\caption{The constraint on the anisotropic BAO dilation parameters $\alpha_{||}$ (top panel), $\alpha_{\perp}$ (middle panel), and the isotropic dilation parameter $\alpha$ (bottom panel). The horizontal and vertical error bars illustrate the width of the redshift bin, and the 68\% CL uncertainty, respectively. The horizontal dashed lines show $\alpha_{||}=\alpha_{\perp}=\alpha=1$ to guide eyes.}
\label{fig:alpha_z}
\end{figure}

\begin{figure}
\centering
{\includegraphics[scale=0.32]{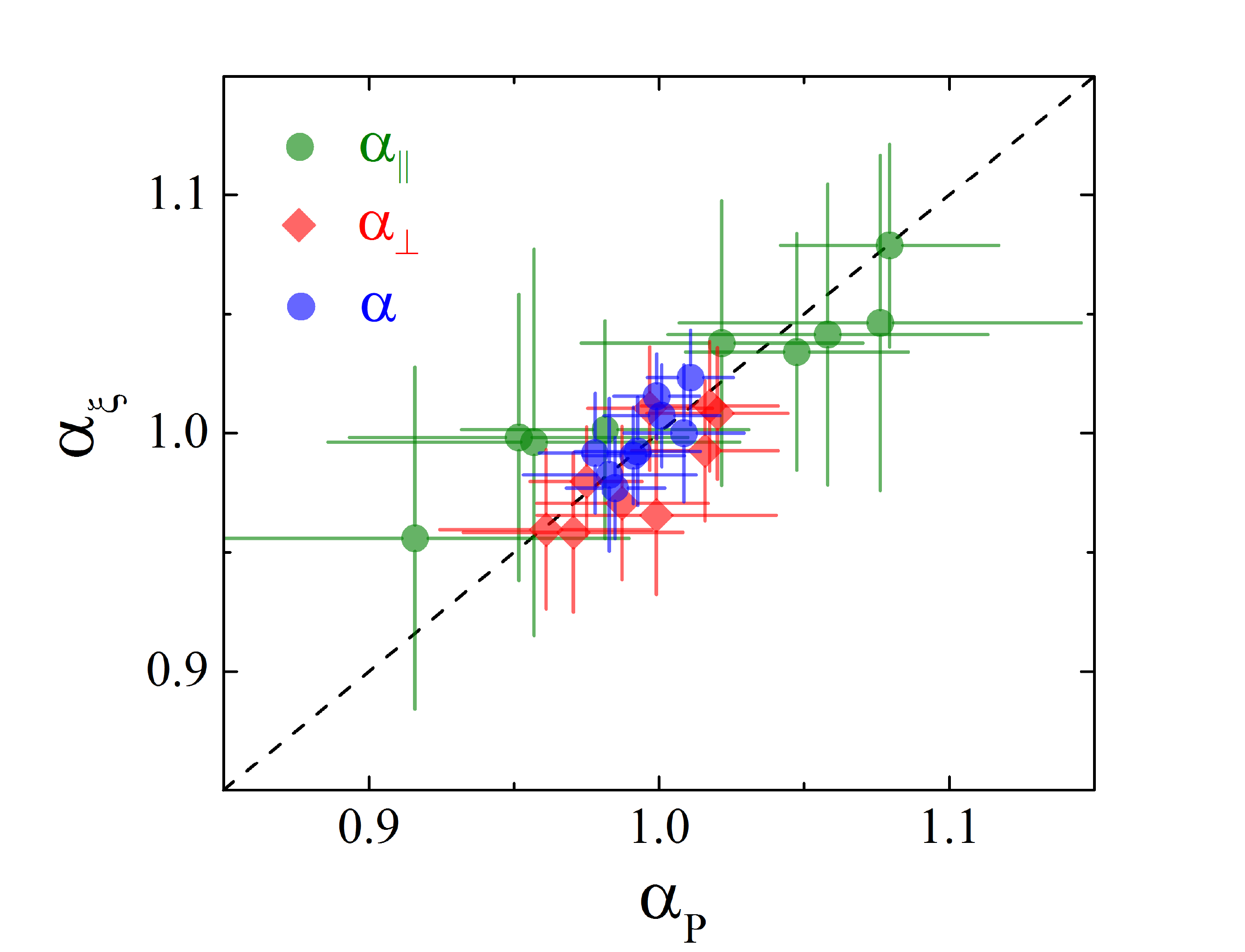}}
\caption{The comparison of our result with that in \citet{tomoBAO-xi}, where $\alpha_{\rm P}$ and $\alpha_{\xi}$ denote the measurements of $\alpha$'s using power spectrum multipoles (this work) and using correlation function multipoles (\citealt{tomoBAO-xi}) respectively.}
\label{fig:alpha_compare}
\end{figure}

\begin{figure}
\centering
{\includegraphics[scale=0.29]{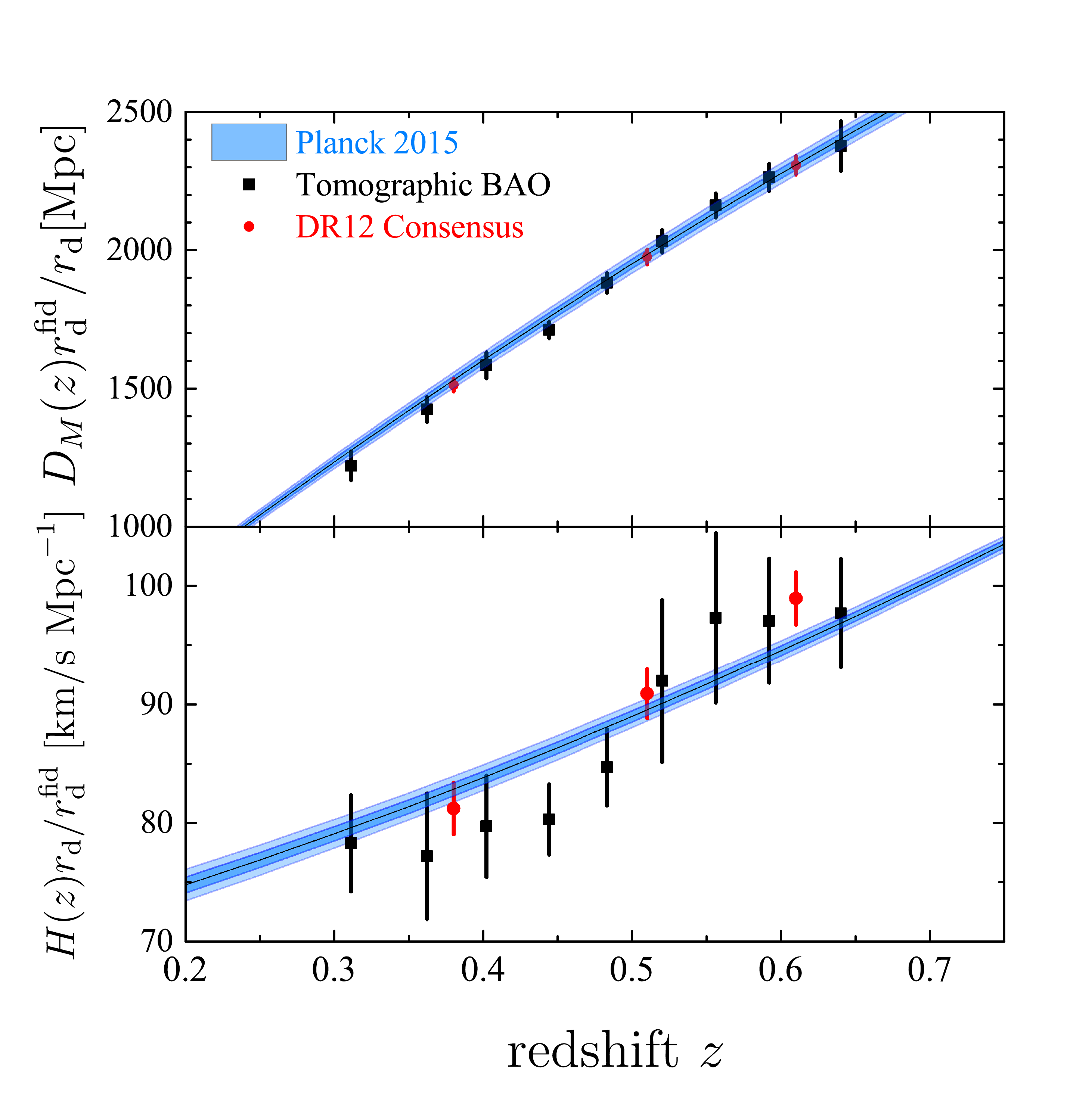}}
\caption{The constraint on $D_M$ and $H$ as a function of redshift, where $D_M\equiv D_A(1+z)$, in comparison with the constraints presented in \citet{Acacia}.}
\label{fig:DM_H}
\end{figure}

Fig \ref{fig:DA_H_gal} shows the contour plots of $D_A/r_d$ and $H \ r_d$, together with the prediction of the fiducial model. Comparing with Fig \ref{fig:DA_H_Fisher}, we find that the degeneracy between $D_A/r_d$ and $H \ r_d$ are consistent with the forecast, while the uncertainties are generally larger, especially for the first and last two bins. This is expected as it is well known that the Fisher forecast, which assumes the Gaussian distribution of parameters, and ignores the systematic effects in the catalogue, can underestimate the errors. Thus we only take the forecast result as a rough guidance for the analysis.

Fig \ref{fig:baoring} visualises the two-dimensional BAO ring in the third redshift slice. The quantity shown in the colours is the two-dimensional power spectrum, which is assembled from our measured $P_0,P_2$ and $P_4$ with the Legendre polynomial, \ie, \be P(k,\mu)=\sum_{\ell=0,2,4} P_{\ell}(k) \mathcal{L}_{\ell}(\mu) \ee To visualise the BAO ring, we divide $P(k,\mu)$ by the smoothed power spectrum $P_{\rm nw}(k,\mu)$.

Fig \ref{fig:alpha_z} shows the constraints on the $\alpha$'s as a function of redshift and in Fig \ref{fig:alpha_compare}, we compare our measurement to the companion paper performing the same tomographic BAO analysis in configuration space \citep{tomoBAO-xi} \footnote{We remove the hexadecapole contribution for this comparison as \citet{tomoBAO-xi} uses the monopole and quadrupole of the correlation function.}. The results are in general consistent with each other within the 68\% CL bound \footnote{Although the correlation function and power spectrum have the same information of BAO in the ideal case (\ie, a survey with an infinite volume without shot noise), a difference is expected for a realistic galaxy survey.}.

The companion paper \citet{Salazar16} performs a similar tomographic BAO analysis, but using different observables and pipeline. \citet{Salazar16} measured the projected two-dimensional angular correlation functions instead in a larger number of redshift slices, and obtained both the BAO and RSD parameters. This method avoids the necessity of choosing a fiducial cosmological model to convert redshifts to distances, which can reduce theoretical systematics in principle, but may be subject to the issue of information loss due to the projection effect, unless a large number of tomographic bins are used \citep{Asorey12}.   

Companion papers \citet{Acacia,Beutler16a,Ross16,Jan16,Sanchez16} perform the BAO measurements using the same galaxy catalogue but in three redshift slices of $0.2<z<0.5, \ 0.4<z<0.6$ and $0.5<z<0.75$. We compare our result to the `DR12 Consensus' result presented in \citet{Acacia} since it is coherently compiled from a range of BAO measurements mentioned above, thus we expect it to be least affected by systematics. 

An overplot of the DR12 Consensus measurement and ours is shown in Fig \ref{fig:DM_H}, with the Planck2015 measurement (mean and 68, 95\% CL errors) shown in blue bands, where $D_M\equiv D_A(1+z)$. A direct one-to-one comparison is impossible simply because our measurements are performed at six additional effective redshifts. The only way for the comparison is to downgrade the redshift resolution of our measurement into three effective redshifts. We follow the procedure presented in \citet{Sanchez16b} for the data compression, and find an agreement within 68\% CL. The comparison is also illustrated in Table 9 and Fig 13 in \citet{Acacia}.

\begin{figure*}
\centering
{\includegraphics[scale=0.35]{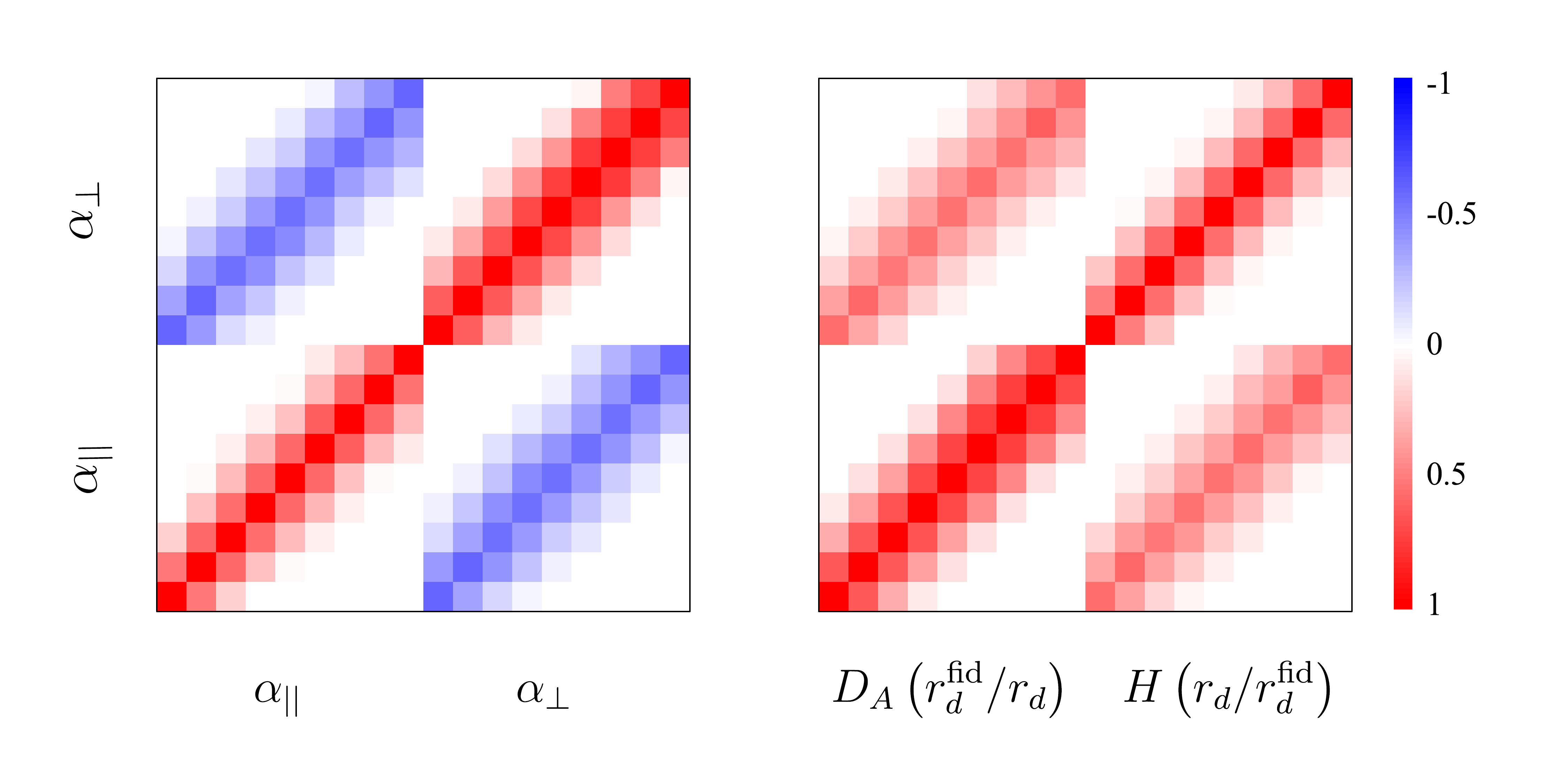}}
\caption{The correlation matrix between $\alpha_{||}$ and $\alpha_{\perp}$ (left) and between $D_A\left(r_d^{\rm fid}/r_d\right)$ and $H\left(r_d/r_d^{\rm fid}\right)$ (right) across all the redshift slices.}
\label{fig:corr_zbins}
\end{figure*}

In order to use our 9-bin tomographic BAO measurement for cosmology, the correlation between redshift bins needs to be quantified. For this purpose, we jointly fit the anisotropic BAO distances in all pairs of overlapping redshift bins, \ie, jointly fit $\alpha_{||}(z_i),\alpha_{||}(z_j),\alpha_{\perp}(z_i),\alpha_{\perp}(z_j)$ with other nuisance parameters marginalised over where $i=1:8; \ j=i+1:9$, and calculate the correlation matrix using the MCMC chain elements. The resultant correlation matrix is shown in Fig \ref{fig:corr_zbins}. As shown, the correlation of the same quantity between redshift bins is positive, and decreases as the redshift separation increases, which is expected. The electronic dataset of measurements presented in this work is available online at \url{https://sdss3.org//science/boss_publications.php}.

\section{Dark energy implications}
\label{sec:modelresult}

In this section, we utilise our tomographic BAO measurements to constrain the equation-of-state (EoS) function of dark energy, $w$, parametrised in the CPL form, \citep{CP, Linder}, \be w(a) = w_0 + w_a(1-a). \ee where $a$ is the scale factor of the Universe. We constrain $w_0,w_a$ together with other basic cosmological parameters including the physical baryon energy density $\Omega_bh^2$, the physical cold dark matter energy density $\Omega_ch^2$, the ratio between the angular diameter distance and sound horizon at recombination $\Theta_s$, the amplitude and power index of the primordial power spectrum $A_s$ and $n_s$ respectively.

Besides the BAO data, we combine with the CMB measurement from the Planck mission \citep{planck15} including the auto- and cross-angular power spectrum of the temperature and polarisation fluctuations of the CMB photons, the supernovae Type Ia sample of JLA \citep{JLA}, the galaxy power spectra from the WiggleZ survey \citep{wigglez}, and the tomographic measurement of the weak lensing shear angular power spectra provided by the CFHTLenS team \citep{cfhtlens}. We pay particular attention to the dark energy perturbations when $w(a)$ crosses the $-1$ boundary \citep{DEP,Fang08}. We use CosmoMC to sample the 7-dimensional parameter space and perform statistical analysis on the Markov chains after the perfect convergence of the sampling process. 

\begin{figure}
\centering
{\includegraphics[scale=0.32]{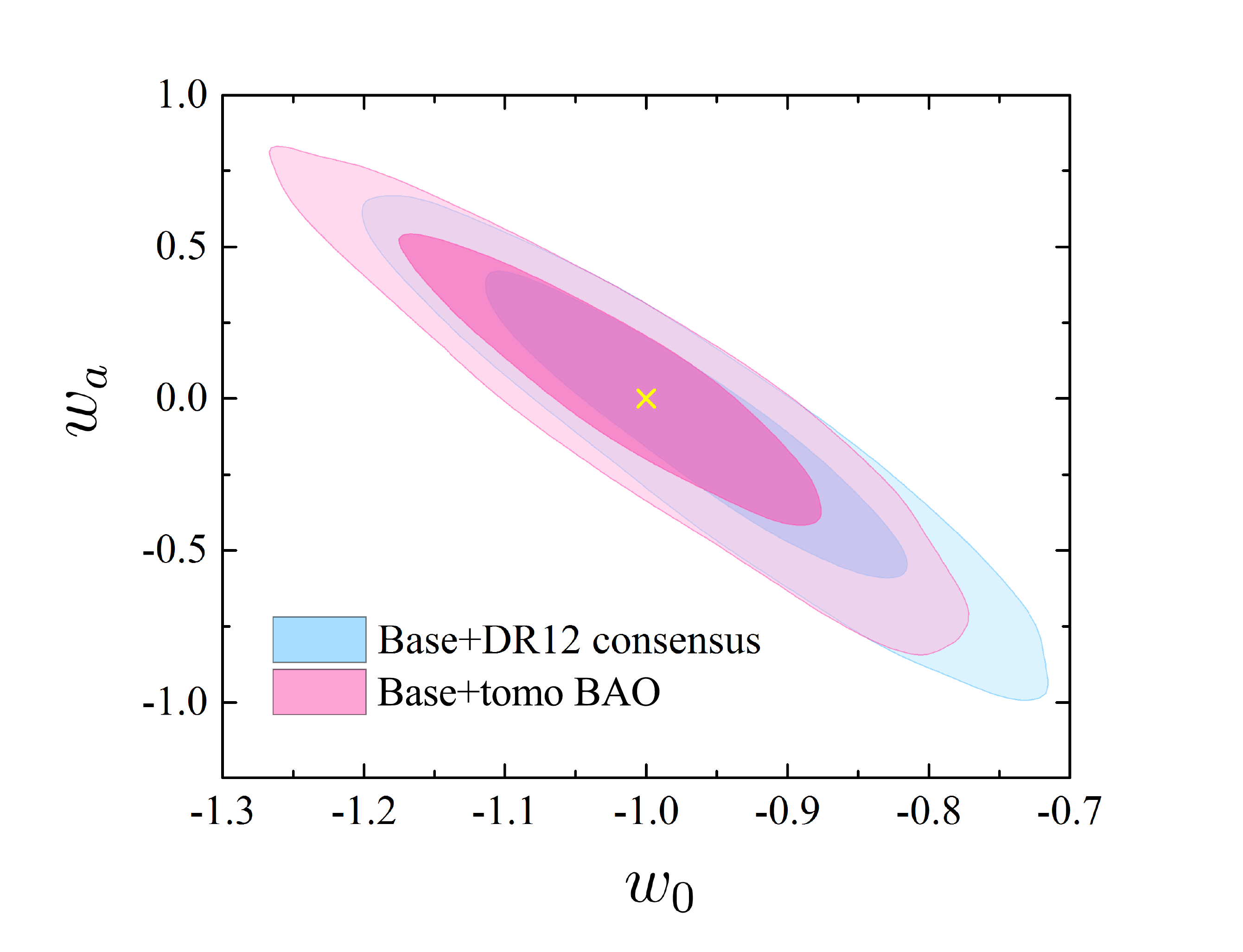}}
\caption{The one-dimensional posterior distribution of $w_0,w_a$ and their two-dimensional 68\% and 95\% CL contour plots derived from the 9-bin tomographic BAO (blue; filled) and the compressed BAO signal at a single redshift (black; unfilled). The Planck 2015 data are combined to complement.}
\label{fig:CPL2}
\end{figure}

The result is visually shown in Fig \ref{fig:CPL2}, where the 68 and 95\% CL contours of $w_0,w_a$ are plotted for two different data combinations (`Base' means a data combination of Planck, JLA, WiggleZ and CFHTLenS). The constraints using our nine-bin tomographic BAO measurements and the three-bin DR12 consensus measurements are consistent well within 68\% CL, while tomographic BAO measurements yield a slightly tighter constraint due to the additional tomographic information in redshift, namely, \ba && w_0 = -0.96\pm 0.10; \ w_a =-0.12\pm0.32 \ ({\rm DR12 \ Consensus}) \nonumber \\ && w_0 = -1.01\pm 0.09; \ w_a =-0.02\pm0.31 \ ({\rm Tomo. \ BAO}) \ea 

\begin{figure}
\centering
{\includegraphics[scale=0.35]{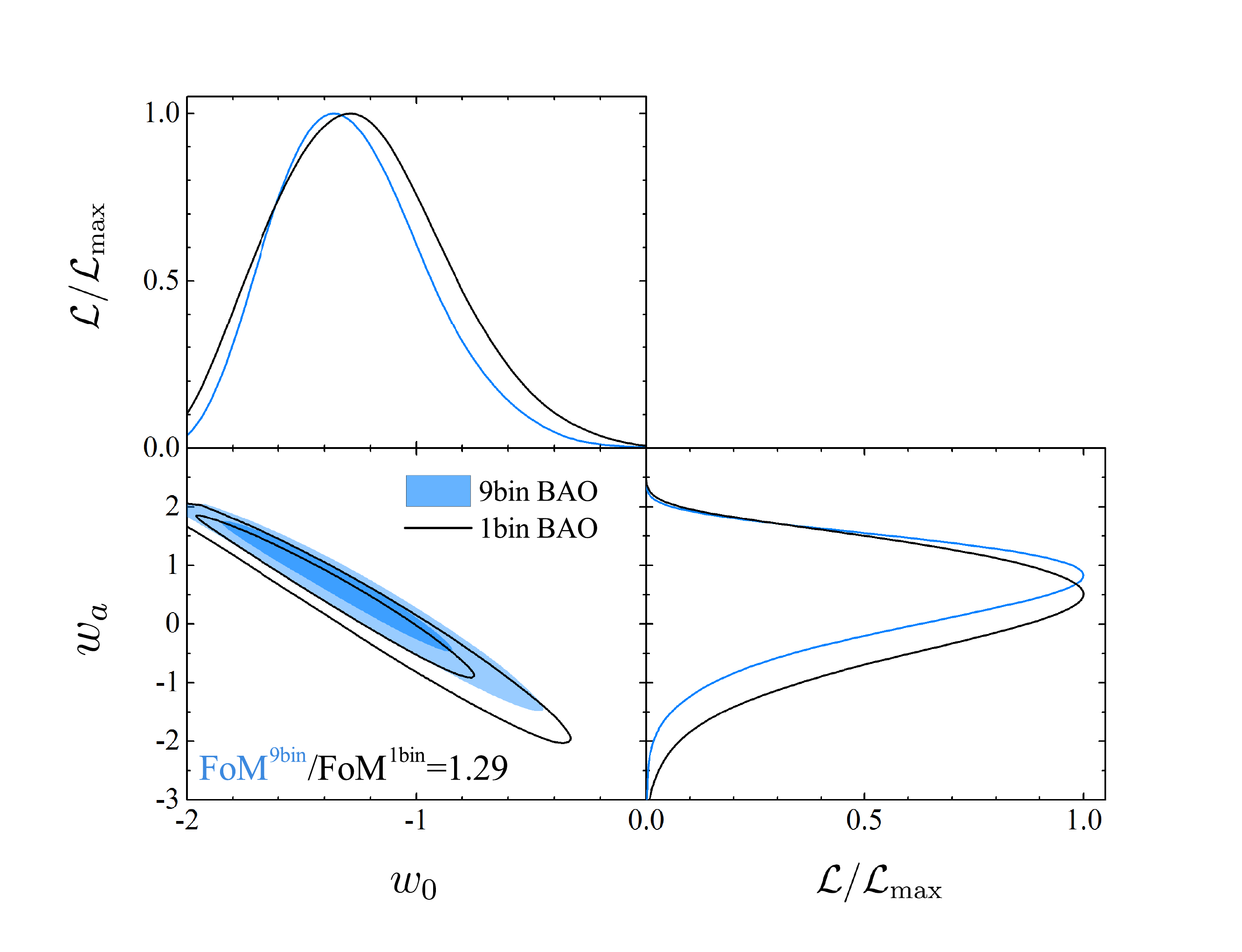}}
\caption{The one-dimensional posterior distribution of $w_0,w_a$ and their two-dimensional 68\% and 95\% CL contour plots derived from the 9-bin tomographic BAO (blue; filled) and the compressed BAO signal at a single redshift (black; unfilled). The Planck 2015 data are combined to complement.}
\label{fig:CPL}
\end{figure}

To quantify the improvement on dark energy parameters using tomographic BAO measurement, we also compare to a test case, in which we maximally remove the tomographic information by compressing our nine-bin BAO measurements into a single datapoint at effective redshift $z_{\rm eff}=0.475$. We also take out the JLA, WiggleZ and CFHTLenS data from the Base dataset to investigate the strength of the DR12 BAO data more explicitly. The result is shown in Fig \ref{fig:CPL}, \ba && w_0 = -1.20\pm 0.32; \ w_a =0.33\pm0.75 \ ({\rm 9 \ bin}) \nonumber \\ && w_0 = -1.18\pm 0.37; \ w_a =0.12\pm0.89 \ ({\rm 1 \ bin })\ea With tomographic BAO, the 68\% CL marginalised errors on $w_0$ and $w_a$ are reduced by 14\% and 16\% respectively, and the Figure-of-Merit (FoM), which is the reciprocal of the area of the 68\% CL $w_0,w_a$ contour, is improved by 29\%. 

Tomographic BAO measurements are informative in terms of the evolution history of $D_A$ and $H$, which are closely related to the time evolution of $w(z)$. It is true that for the CPL parametrisation, the improvement from the current tomographic BAO measurement is not significant, but the tomographic BAO is much more informative for the non-parametric reconstruction of $w(z)$ \citep{Zhao16}. Moreover, for future galaxy surveys which cover a wider redshift range, the improvement on $w(z)$ constraint is expected to be more significant. 

\section{Conclusion}
\label{sec:conclusion}

The physics of baryonic acoustic oscillations has been well established to be a robust tool for cosmological studies. Specifically, the BAO measurements make it possible to reconstruct the history of the cosmic expansion, which is key to revealing the physics of the accelerating expansion of the Universe, and the nature of dark energy. 

Obtaining BAO measurements at as many redshifts as possible is ideal for tracing the cosmic expansion history. However, extracting the time evolution of the BAO signal is technically challenging. Na\"{\i}vely subdividing the galaxies into multiple independent redshift slices and performing BAO measurements in each slice is a straightforward solution, but the number of slices has to be limited to a small number, otherwise each individual slice would contain too few galaxies to enable a robust BAO measurement due to the low signal-to-noise ratio and issues of systematics. 

In this work, we solve this problem using multiple {\it overlapping} redshift slices, which allows for extracting the redshift information of the BAO signal in a large number of redshift slices. We exploit the completed DR12 combined galaxy sample of the BOSS survey, and obtain tomographic BAO measurements in nine overlapping redshift slices using the pre-reconstructed galaxy power spectrum multipoles up to the hexadecapole, after validating our data analysis pipeline using the MD-Patchy mock galaxy catalogues. Our measurement and likelihood routines compatible with CosmoMC are publicly available. 

We compare our measurement to that in a companion paper \citep{tomoBAO-xi}, which performs similar analysis using galaxy correlation functions derived from the same data sample, and find consistent results. For a further comparison, we derive a three-bin BAO measurement by coherently combining our tomographic measurements, and then compare to the BAO measurement presented in another companion paper \citep{Acacia}, and find an agreement \footnote{Note that, besides the different redshift binning scheme from that used in \citet{Acacia}, this work differs in two aspects: we use the fourth-order B-spline to obtain the overdensity field on the grid, which largely removes the aliasing effect; and include the hexadecapole in the BAO analysis, which we find indeed helps with the BAO constraint.}. The BAO measurements including the full covariance matrices presented in this work and a {\tt CosmoMC} patch is available at \url{https://sdss3.org//science/boss_publications.php}.

We use our BAO measurements to constrain dark energy equation-of-state parameters, and find that for the CPL parametrisation, the $\Lambda$CDM model is favoured by a joint dataset of CMB, supernovae, BAO and weak lensing measurement. A more generic approach for dark energy studies using our measurement will be explored in a separate publication \citep{Zhao16}.

For the BOSS DR12 sensitivity, we have seen that the dark energy FoM can differ by as much as 29\% between cases using tomographic, and non-tomographic BAO measurements. The ongoing and upcoming galaxy redshift surveys, including the eBOSS \footnote{\url{http://www.sdss.org/surveys/eboss/}} \citep{Overview}, DESI \footnote{\url{http://desi.lbl.gov/}}, Euclid \footnote{\url{http://www.euclid-ec.org/}} \citep{Euclid16}, PFS \footnote{\url{http://sumire.ipmu.jp/pfs/}} \citep{PFS}, and so on, cover a larger and larger cosmic volume, thus there is rich tomographic information in redshifts to be exploited. Besides the method developed in this work, alternatives such as the optimal redshift weighting method \citep{zweight1,zweight2,zweight3}, are being developed and applied to galaxy surveys. 
       
\section*{Acknowledgements}

GBZ and YW are supported by the Strategic Priority
Research Program ``The Emergence of Cosmological
Structures" of the Chinese Academy of Sciences Grant
No. XDB09000000, by National Astronomical Observatories, Chinese Academy of Sciences (NAOC), and by University of Portsmouth. GBZ is supported by the 1000 Young Talents program in China. YW is supported by the NSFC grant No. 11403034. G.R. is supported by the National Research Foundation of Korea (NRF) through NRF-SGER 2014055950 funded by the Korean Ministry of Education, Science and Technology (MoEST), and by the faculty research fund of Sejong University in 2016. 

Funding for SDSS-III has been provided by the Alfred
P. Sloan Foundation, the Participating Institutions, the
National Science Foundation, and the U.S. Department
of Energy Office of Science. The SDSS-III web site is
\url{http://www.sdss3.org/}.
SDSS-III is managed by the Astrophysical Research
Consortium for the Participating Institutions of the SDSS-
III Collaboration including the University of Arizona, the
Brazilian Participation Group, Brookhaven National Laboratory, Carnegie Mellon University, University of Florida,
the French Participation Group, the German Participation
Group, Harvard University, the Instituto de Astrosica de Canarias, the Michigan State/Notre Dame/JINA Participation Group, Johns Hopkins University, Lawrence Berkeley National Laboratory, Max Planck Institute for Astro-
physics, Max Planck Institute for Extraterrestrial Physics,
New Mexico State University, New York University, Ohio
State University, Pennsylvania State University, University
of Portsmouth, Princeton University, the Spanish Participation Group, University of Tokyo, University of Utah,
Vanderbilt University, University of Virginia, University of
Washington, and Yale University.

This research used resources of the National Energy Re-
search Scientific Computing Center, which is supported by
the Office of Science of the U.S. Department of Energy under
Contract No. DE-AC02-05CH11231, the SCIAMA cluster supported by University of Portsmouth, and the ZEN cluster supported by NAOC.  

\bibliographystyle{mn2e}
\bibliography{tomoBAOpk}

\label{lastpage}

\end{document}